\documentclass[showpacs,preprintnumbers,nofootinbib,
superscriptaddress,amsmath,floatfix,prd]{revtex4}

\usepackage{amssymb}  
\usepackage{graphicx,subfigure}
\usepackage{exscale}
\usepackage{textcomp}
\usepackage{enumerate}

\usepackage{color}

\usepackage[normalem]{ulem}  

\newcommand{\non}{\nonumber\\}

\newcommand{\be}{\begin{equation}}
\newcommand{\ee}{\end{equation}}
\newcommand{\bea}{\begin{eqnarray}}
\newcommand{\eea}{\end{eqnarray}}
\newcommand{\ba}[1]{\begin{array}{#1}}
\newcommand{\ea}{\end{array}}

\newcommand{\bm}[1]{\mbox{\boldmath${#1}$}}

\begin{document}

\title{Magnetic catalysis in nuclear matter}

\author{Alexander Haber}
\email{ahaber@hep.itp.tuwien.ac.at}
\affiliation{Institut f\"{u}r Theoretische Physik, Technische Universit\"{a}t Wien, 1040 Vienna, Austria}

\author{Florian Preis}
\email{fpreis@hep.itp.tuwien.ac.at}
\affiliation{Institut f\"{u}r Theoretische Physik, Technische Universit\"{a}t Wien, 1040 Vienna, Austria}

\author{Andreas Schmitt}
\email{aschmitt@hep.itp.tuwien.ac.at}
\affiliation{Institut f\"{u}r Theoretische Physik, Technische Universit\"{a}t Wien, 1040 Vienna, Austria}

\date{10 December 2014}

\begin{abstract}

A strong magnetic field enhances the chiral condensate at low temperatures. This so-called magnetic catalysis thus 
seeks to increase the 
vacuum mass of nucleons. We employ two relativistic field-theoretical models for nuclear matter, the Walecka model and an extended linear sigma model, to discuss the 
resulting effect on the transition between vacuum and nuclear matter at zero temperature. 
In both models we find that the creation of nuclear matter in a sufficiently strong magnetic field becomes energetically more costly due to the 
heaviness of magnetized nucleons, even though it is also found that nuclear matter is more strongly bound in a magnetic field.
Our results are potentially important for dense nuclear matter in compact stars, especially since previous studies in the astrophysical 
context have always ignored the contribution of the magnetized Dirac sea and thus the effect of magnetic catalysis.

\end{abstract}


\maketitle


\section{Introduction}

The chiral condensate of Quantum Chromodynamics (QCD) is, at sufficiently low temperatures, enhanced by a background magnetic field 
\cite{D'Elia:2010nq,D'Elia:2011zu,Bali:2011qj,Bali:2012zg} due to magnetic catalysis 
\cite{Klimenko:1990rh,Klimenko:1992ch,Gusynin:1994re,Gusynin:1994va,Gusynin:1994xp,Shushpanov:1997sf,Agasian:2001hv,Miransky:2002rp,Shovkovy:2012zn,Kharzeev:2013jha}. 
At weak coupling, magnetic catalysis 
is analogous to Cooper pairing \`{a} la Bardeen-Cooper-Schrieffer (BCS): in both cases, an effective dimensional reduction leads to the creation of 
a condensate. In BCS theory, it is the Fermi surface that renders the gap equation effectively 1+1 dimensional and induces a fermion-fermion
condensate for an arbitrarily weak attractive interaction. In the case of magnetic catalysis, a sufficiently strong magnetic field plays a similar role by suppressing the 
dynamics in the directions perpendicular to the magnetic field. As a consequence, a fermion-antifermion condensate is induced, also for arbitrarily weak 
attractive interactions. 
In the QCD vacuum, the coupling is strong of course, and a chiral condensate is present even without magnetic field. Nevertheless, a magnetic field 
further increases this condensate at low temperatures\footnote{One might thus expect the critical temperature for chiral symmetry breaking in QCD to increase
with the magnetic field. However, at larger temperatures, the situation is more complicated, and the critical temperature appears to 
{\it decrease} \cite{Bali:2011qj}. Such an ``inverse magnetic catalysis''  
has also been discussed in holographic and field-theoretical models at large baryon chemical potential and low temperatures (where it is of completely different 
physical origin) \cite{Preis:2010cq,Preis:2012fh}.}.

In this paper, we discuss the effect of an external magnetic field $B$ on nuclear matter, in particular on the transition between the vacuum and nuclear 
matter at zero temperature. In the absence of a magnetic field, we know that this onset is a first-order phase transition and that it occurs 
at a baryon chemical potential $\mu\simeq 923\,{\rm MeV}$ which is smaller than the vacuum mass of the nucleon. 
The reason is that the energy per baryon is reduced by the binding energy of nuclear matter. A strong magnetic field of the 
order of or larger than the QCD scale, $\Lambda_{\rm QCD}^2 \sim 10^{18}\,{\rm G}$, can be expected to affect both, the vacuum mass and the binding energy. 
Therefore, even if the vacuum mass is enhanced by magnetic catalysis, it is not a priori clear whether the critical chemical 
potential for the onset is enhanced too. In fact, we shall see that the critical chemical potential is a non-monotonic function of the magnetic field. 

We perform two separate calculations within two models for nuclear matter. Firstly, we use the Walecka model \cite{Walecka:1974qa,Serot:1984ey}, including
scalar self-interactions \cite{Boguta:1977xi}, where the 
interaction between nucleons is modelled by the exchange of $\sigma$ and $\omega$ mesons. Secondly, we employ an extended linear sigma model, including 
nucleons and their chiral partners \cite{Detar:1988kn,Jido:1998av,Jido:2001nt,Zschiesche:2006zj,Gallas:2009qp,Gallas:2011qp,Heinz:2013hza}. 
Both models contain various parameters which are fitted to reproduce vacuum masses of 
mesons and nucleons as well as properties of nuclear matter at the saturation density in the absence of a magnetic field. 
One important difference between the two models is 
the origin of the vacuum mass of the nucleons: while in the Walecka model it is a given parameter, 
in the extended linear sigma model used here it is generated dynamically by spontaneous breaking of chiral symmetry. 

Nuclear matter is subject to large magnetic fields in the extreme environment of compact stars. 
Surface magnetic fields of compact stars (then called magnetars)
can be as large as $10^{15}\,{\rm G}$ \cite{Duncan:1992hi} (for a review, see Ref.\ \cite{2011heep.conf..247R}). It is conceivable -- although speculative --
that in the interior of magnetars the magnetic field might be several orders of magnitude larger and thus affect dense matter on the scale of QCD \cite{Lai}. 
Besides the static properties of magnetars, magnetic fields may also play a prominent role in compact star mergers. 
The gravitational waves emitted in the late
stage of such a merger process can potentially be observed directly and are  sensitive to the equation of state of nuclear matter 
\cite{Read:2013zra}. The magnetic field in a merger process might become extremely large through a magneto-rotational 
instability \cite{Siegel:2013nrw}, such that corrections to the equation of state due to a magnetic field may become important.

Dense nuclear matter in a background magnetic field has been studied before, using a relativistic mean-field approach with interactions
through $\sigma$, $\omega$, and $\rho$ 
mesons \cite{2000ApJ...537..351B,Broderick:2001qw,Sinha:2010fm,Rabhi:2011ej,Dexheimer:2011pz,Preis:2011sp,Dong:2013hta,deLima:2013dda,Casali:2013jka}. 
In all these works, the (divergent) vacuum contribution was omitted\footnote{Including the work coauthored by two of the present authors \cite{Preis:2011sp}, 
whose main calculation made use of the holographic 
Sakai-Sugimoto model. A field-theoretical mean-field study was used for comparison and found to be in disagreement with the holographic result. 
The results of the present paper
show that the disagreement was partly due to the missing vacuum contribution.}. 
In the absence of a magnetic field, it has been shown that this ``no-sea approximation'' only leads to a very small 
difference in the equation of state compared to the result where this contribution is kept and the theory properly 
renormalized \cite{1988PhLB..208..335G,1989NuPhA.493..521G}. However, magnetic catalysis occurs in the vacuum. 
Therefore, throwing out the vacuum contribution amounts to throwing out important physics, and at least the $B$-dependent part should
be taken into account carefully (we shall show that the $B$-independent part remains negligible in our results). 
This has been done in the original works about magnetic catalysis as well as in many following studies, 
for instance in the 
Nambu--Jona-Lasinio (NJL) model \cite{Ebert:1999ht,Inagaki:2003yi,Menezes:2008qt,Preis:2012fh,Gatto:2012sp,Ferrari:2012yw}, a quark-meson 
model \cite{Ferrari:2012yw,Andersen:2011ip,Skokov:2011ib,Andersen:2012bq}, and the MIT bag model \cite{Fraga:2012fs}. 

The present paper is, to our knowledge,  the first to include the effect of magnetic catalysis in a relativistic mean-field description of nuclear matter.
We shall concentrate on the onset of nuclear matter, where the effects of the magnetic field can be explained in a very transparent way.  
We do not attempt to make any quantitative predictions for the role of magnetic catalysis for matter in the interior of compact stars, which can be several times 
denser than the matter we discuss here. Moreover, we work with a very simple version of nuclear matter. We consider isospin-symmetric matter 
(the only isospin-breaking 
effect coming from the different electric charges of neutrons and protons), neglect the anomalous magnetic moments, and do not require our
matter to be electrically neutral or in chemical equilibrium. Also, we will not take into account superfluidity of the nucleons.
Our study is therefore a starting point for more realistic calculations -- or, in other words, it should be used for improving 
existing studies of dense nuclear matter in a magnetic field. 

The paper is organized as follows. In Sec.\ \ref{sec:free} we introduce the free energy in a form that is valid for both models we consider. The renormalization
of this free energy is discussed in Sec.\ \ref{sec:renorm}, and in Sec.\ \ref{sec:models} we introduce the two models in detail. The results of 
our calculations are presented in Secs.\ \ref{sec:vacmasses} and \ref{sec:onset}: in Sec.\ \ref{sec:vacmasses} we compute the vacuum masses of the nucleons
as a function of the magnetic field, and in Sec.\ \ref{sec:onset} we present the zero-temperature onset of nuclear matter in the presence of 
a magnetic field. We give our conclusions in Sec.\ \ref{sec:summary}.

\section{Free energy}
\label{sec:free}

In both models we consider, the unrenormalized free energy density can be written as
\be \label{Omgen}
\Omega = \frac{B^2}{2} + U + \Omega_{N} \, , 
\ee
where $B^2/2$ is the field energy of the magnetic field, which, without loss of generality, points in the $z$-direction, ${\bf B}=(0,0,B)$, 
and $U$ is the tree-level potential that is independent of the nucleons. It will be specified for the two models separately in Sec.\ \ref{sec:models}; 
its explicit form does not play any role now and for the renormalization discussed in Sec.\ \ref{sec:renorm}.
The nucleonic part $ \Omega_{N} $ depends on the mass $M$ of the nucleons (several baryon masses $M_i$ in general) 
and the externally given thermodynamic parameters $B$, baryon chemical potential $\mu$, and temperature $T$.
Since $M$ will be determined dynamically by minimizing $\Omega$, it depends on $B$, $\mu$, and $T$ implicitly. 
Therefore, we can write $\Omega_{N}  = \Omega_N[M(B,\mu,T),B,\mu,T]$, and decompose
\be \label{decompose}
\Omega_N = \Omega_{N,{\rm sea}} + \Omega_{N,{\rm mat}} \, ,
\ee
where $\Omega_{N,{\rm sea}}  \equiv \Omega_N[M(B,\mu,T),B,0,0]$ is the free energy of the magnetized vacuum. 
Because of the medium dependence of the nucleon mass, $\Omega_{N,{\rm sea}}$ is not a vacuum 
contribution in the strict sense. We shall thus mostly refer to it as the contribution of the Dirac sea or, briefly, the ``sea contribution''.
It depends on the ultraviolet cutoff, and we discuss its renormalization in the next section, while the matter contribution $\Omega_{N,{\rm mat}}$ is finite.
(Had we separated the ``pure'' magnetized vacuum $\Omega_N[M(B,0,0),B,0,0]$, the remaining matter part would not have been finite.)

In the mean-field approximation, $\Omega_N$ assumes the form of free fermions, with all interaction effects absorbed in the 
medium-dependent nucleon mass and an effective baryon chemical potential $\mu_*$. 
In the vacuum, i.e., for $T=0$ and $\mu_*<M$, we have $\mu=\mu_*$. 
For a spin-$\frac{1}{2}$ fermion with (bare) electric charge $q$ and mass $M$, the sea contribution is 
\be    \label{OmNvac}
\Omega_{N,{\rm sea}} = - \frac{|qB|}{2\pi} \sum_{\nu=0}^\infty \alpha_\nu\int_{-\infty}^\infty\frac{dk_z}{2\pi}\epsilon_{k,\nu} \, , 
\ee
where
\be \label{epskn}
\epsilon_{k,\nu} = \sqrt{k_z^2 + 2\nu|qB| + M^2} 
\ee
are the single-fermion energies, and where the sum over $\nu$ refers to the Landau levels. The factor $\alpha_\nu\equiv 2-\delta_{\nu 0}$ accounts 
for the spin degeneracy of each Landau level (only fermions with a single spin polarization occupy the lowest Landau level $\nu=0$). 
The matter part is given by 
\bea \label{OmNmat}
\Omega_{N,{\rm mat}} &=&  - \frac{|qB|T}{2\pi}\sum_{e=\pm} \sum_{\nu=0}^\infty \alpha_\nu\int_{-\infty}^\infty\frac{dk_z}{2\pi} 
\ln\left(1+e^{-\frac{\epsilon_{k,\nu}-e \mu_*}{T}}\right) \non[2ex]
&\stackrel{T=0}{\longrightarrow}& 
- \frac{|qB|}{4\pi^2} \Theta(\mu_*-M) \sum_{\nu=0}^{\nu_{\rm max}} \alpha_\nu\left[\mu_*k_{F,\nu}-(M^2+2\nu|qB|)\ln\frac{\mu_*+k_{F,\nu}}{\sqrt{M^2+2\nu|qB|}}\right]
\, ,
\eea
where $e=-1$ corresponds to the anti-particle contribution which disappears at $T=0$ because of $\mu_*>0$.
In the zero-temperature expression, we have defined the Fermi momentum in the $z$-direction for each Landau level,
\be
k_{F,\nu}\equiv \sqrt{\mu_*^2-(M^2+2\nu |qB|)} \, , 
\ee 
and the upper limit for the sum over Landau levels,
\be
\nu_{\rm max} \equiv \left\lfloor \frac{\mu_*^2-M^2}{2|qB|}\right\rfloor \, . 
\ee 
For neutral fermions, $q=0$, we have  
\be \label{OmNvac0}
\Omega_{N,{\rm sea}}(qB=0) = - 2 \int\frac{d^3{\bf k}}{(2\pi)^3}\epsilon_{k} \, ,  
\ee
and 
\bea \label{OmNmat0}
\Omega_{N,{\rm mat}}(qB=0) &=&  - 2T\sum_{e=\pm} \int\frac{d^3{\bf k}}{(2\pi)^3}\ln\left(1+e^{-\frac{\epsilon_{k}-e \mu_*}{T}}\right) \non[2ex]
&\stackrel{T=0}{\longrightarrow}& -\frac{\Theta(\mu_*-M)}{8\pi^2}\left[\left(\frac{2}{3}k_F^3-M^2k_F\right)\mu_*+M^4\ln\frac{k_F+\mu_*}{M}\right]   \, ,
\eea
with the excitation energy
\be
\epsilon_k=\sqrt{k^2+M^2} \, , 
\ee
and the Fermi momentum 
\be
k_F \equiv \sqrt{\mu_*^2-M^2} \, .
\ee 
We shall use this free energy to describe neutrons, while the above expressions for charged fermions shall be used for the protons. 
We neglect the anomalous magnetic moment for the sake of simplicity, 
which in particular means that the neutrons are not affected by the magnetic field at all. In previous studies, 
the anomalous magnetic moment has mostly been included within an effective approach \cite{2000ApJ...537..351B,Broderick:2001qw,Sinha:2010fm,Rabhi:2011ej,Dexheimer:2011pz,Preis:2011sp,Dong:2013hta,deLima:2013dda,Casali:2013jka}. This approach is valid for not-too-large magnetic fields. We leave a more realistic study,
including the anomalous magnetic moments, for the future. Because of the importance of the renormalization of the sea terms, 
it remains to be seen whether the widely used effective approach is appropriate or whether a more microscopic approach, 
for instance along the lines of Ref.\ \cite{Paret:2014hxa}, should be considered for generalizing the renormalization that we discuss now.

\section{Renormalization}
\label{sec:renorm}

There are many works in the literature that have discussed the renormalization of the free energy of charged fermions in 
a magnetic field. Even though this renormalization has, to our knowledge, never been applied to a relativistic mean-field model for nuclear matter, 
we can proceed exactly as for instance in the NJL model. Therefore, our main point is not the derivation of the renormalized free energy, but its
application to nuclear matter. Nevertheless, we shall go through the renormalization procedure in some detail. The reason is that 
there exist different results for the free energy after renormalization in the literature, and we will point out that these results correspond to different 
choices of the renormalization scale. 

We consider the free energy for charged fermions. Thus, we need to regularize the divergent integral in Eq.\ (\ref{OmNvac}), which we do with the help of the 
proper time method \cite{Schwinger:1951nm} that has been widely used in the related literature, see for instance 
Refs.\ \cite{Gusynin:1994re,Gusynin:1994va,Gusynin:1994xp,Inagaki:2003yi,Preis:2012fh} 
(dimensional regularization leads to the same result \cite{Menezes:2008qt,Andersen:2011ip,Fraga:2012fs,Endrodi:2013cs}). 
By rewriting the integrand $\epsilon_{k,\nu}$ 
with the help of   
\be
\frac{1}{x^a} = \frac{1}{\Gamma(a)}\int_0^\infty d\tau\,\tau^{a-1} e^{-\tau x} \, ,
\ee
performing the momentum integral and the sum over all Landau levels,
we find
\bea
\Omega_{N,{\rm sea}} &=& \frac{|qB|}{8\pi^2}\int_0^\infty\frac{d\tau}{\tau^2}e^{-\tau M^2}\coth(|qB|\tau) \, . 
\eea
This integral is still divergent, and we replace the lower boundary by $1/\Lambda^2$, such that the result depends on the ultraviolet cutoff $\Lambda$. 
In the limit of large $\Lambda$ we obtain
\bea \label{resultOmNvac}
\Omega_{N,{\rm sea}} &=& \Omega_{N,{\rm sea}}(qB=0) -  \frac{|qB|^2}{24\pi^2}\left(\gamma + \ln\frac{M^2}{\Lambda^2}\right) \non[2ex]
&& -\frac{|qB|^2}{2\pi^2} \left[\frac{x^2}{4}(3-2\ln x) +\frac{x}{2}\left(\ln\frac{x}{2\pi}-1\right)+\psi^{(-2)}(x)- \frac{\ln A^{12}x}{12} \right] \, , 
\eea
where $\gamma \simeq 0.577$ is the Euler-Mascheroni constant, $\psi^{(n)}$ the $n$-th polygamma function (analytically continued to negative $n$), 
and $A \simeq 1.282$ the Glaisher constant [$\ln A = \frac{1}{12}-\zeta'(-1)$, 
with the Riemann zeta function $\zeta$]. We have abbreviated 
\be \label{x}
x\equiv \frac{M^2}{2|qB|} \, , 
\ee
and we have separated the contribution of the unmagnetized Dirac sea (\ref{OmNvac0}), which, applying the same proper time regularization, reads  
\bea \label{OmegaBeq0}
\Omega_{N,{\rm sea}}(qB=0) &=&  \frac{1}{16\pi^2}\left[\Lambda^2(\Lambda^2-M^2)e^{-M^2/\Lambda^2}+M^4\Gamma\left(0,\frac{M^2}{\Lambda^2}\right) 
\right] \, , 
\eea 
where $\Gamma(a,x)$ is the incomplete gamma function.

In the absence of a magnetic field, usually the ``no-sea approximation'' is employed in the context of mean-field models for nuclear matter, i.e., 
$\Omega_{N,{\rm sea}}$ is ignored. Since $\Omega_{N,{\rm sea}}$ depends implicitly on the medium, and it thus contributes to the minimization of the free energy in a 
non-trivial way, there is a priori no reason for this approximation to be valid. It has been justified by an explicit check of
the smallness of its correction to the final result for the equation of state  
\cite{1988PhLB..208..335G,1989NuPhA.493..521G}. Here we distinguish between $\Omega_{N,{\rm sea}}(qB=0)$ and the contribution of the magnetized Dirac sea. 
In appendix \ref{AppA} we show, for the case of the Walecka model, that $\Omega_{N,{\rm sea}}(qB=0)$ has a very small effect on our results, 
and we will thus proceed solely with the $B$-dependent sea contribution. To illustrate the qualitative difference between the two
contributions, a comparison with the NJL model -- whose degrees of freedom are quarks, not nucleons -- is instructive: 
in the NJL model, the $B$-{\it independent} sea contribution is responsible for chiral symmetry breaking 
in the vacuum for coupling strengths larger than a critical coupling, and it clearly must not be discarded, even though it introduces a cutoff dependence
in the non-renormalizable NJL model. The $B$-{\it dependent} sea contribution is responsible for 
magnetic catalysis in the vacuum, inducing a chiral condensate for arbitrarily small coupling strength. Now, in our present study of nuclear matter, chiral 
symmetry breaking in the vacuum is, in the Walecka model, put in by hand through a given vacuum mass of the nucleons and, in the extended linear sigma model, generated 
dynamically by chiral symmetry breaking. 
Therefore, dropping the $B$-{\it independent} sea contribution does not throw out important physics, and we only have to check 
whether its quantitative effect is small in our results, see appendix \ref{AppA}. In contrast, dropping the $B$-{\it dependent} sea contribution, 
as done in Refs.\ \cite{2000ApJ...537..351B,Broderick:2001qw,Sinha:2010fm,Rabhi:2011ej,Dexheimer:2011pz,Preis:2011sp,Dong:2013hta,deLima:2013dda,Casali:2013jka}, does throw out
important physics, namely magnetic catalysis. Hence we keep it. 

The expression on the right-hand side of Eq.\ (\ref{resultOmNvac}) contains a logarithmic cutoff dependence. This dependence can be absorbed into a renormalized 
magnetic field and a renormalized electric charge. To this end, we introduce the renormalized charge by $q^2 = Z_q^{-1} q_r^2$ and the renormalized 
magnetic field by $B^2=Z_q B_r^2$ such that $qB=q_rB_r$, where 
\be
Z_q = 1+ \frac{q_r^2}{12\pi^2}\left(\gamma+\ln\frac{\ell^2}{\Lambda^2} \right) \, ,
\ee
with a  renormalization scale $\ell$. We can thus write 
\bea \label{Omrenorm}
\frac{B^2}{2} + \Omega_{N,{\rm sea}} &=& \frac{B_r^2}{2}  - \frac{|q_rB_r|^2}{24\pi^2}\ln\frac{M^2}{\ell^2} 
- \frac{|q_rB_r|^2}{2\pi^2}\left[\frac{x^2}{4}(3-2\ln x) +\frac{x}{2}\left(\ln\frac{x}{2\pi}-1\right)+\psi^{(-2)}(x) - \frac{\ln A^{12}x}{12} \right] \non[2ex]
&=& \frac{B_r^2}{2} - \frac{|q_rB_r|^2}{24\pi^2}\ln\frac{2|q_rB_r|}{\ell^2A^{12}} - \frac{|q_rB_r|^2}{2\pi^2}\left[\frac{x^2}{4}(3-2\ln x) 
+\frac{x}{2}\left(\ln\frac{x}{2\pi}-1\right)+\psi^{(-2)}(x)\right]\, ,
\eea
where only renormalized quantities appear. 
We have written the result in two different ways to make the discussion about the choice of the renormalization scale $\ell$ more transparent. There seem to be 
two natural choices for $\ell$. If we choose the nucleon mass, $\ell = M$, we read off the nonvanishing terms in the first line, while, if we choose the 
magnetic field as a scale, $\ell = \sqrt{2|q_rB_r|}/A^6 \simeq 0.318 \sqrt{|q_rB_r|}$, the second line shows that this choice corresponds to 
keeping only terms that depend on $M$ (plus the free field term)\footnote{Also for $n$, instead of 1, charged nucleonic states, the logarithms vanish for 
particular choices of the renormalization scale: if we choose the mass scale,
$\ell = (M_1^{p_1^2}M_2^{p_2^2}\ldots M_n^{p_n^2})^{1/(p_1^2+p_2^2+\ldots + p_n^2)}$, and if we choose the magnetic field as a scale,
$\ell = \sqrt{2|e_rB_r|}/A^6$ $\times(|p_1|^{p_1^2/2}|p_2|^{p_2^2/2}\ldots|p_n|^{p_n^2/2})^{1/(p_1^2+p_2^2+\ldots + p_n^2)}$.
Here, $M_i$ is the mass of the $i$-th nucleon and $p_i = q_i/e$ its charge in units of the elementary charge $e$.}. 

The choice for $\ell$ matters for evaluating observables such as the 
magnetization or the pressure itself. It has been pointed out in Ref.\ \cite{Endrodi:2013cs} (see also Ref.\ \cite{Dunne:2004nc}), that only for 
$\ell = M$ the vacuum pressure for small magnetic fields $x\gg 1$ is proportional to $B_r^2$, receiving its sole contribution from the free field 
term because all other contributions are of order $B_r^4$ and higher, 
\bea
x\gg 1: \qquad \frac{x^2}{4}(3-2\ln x) +\frac{x}{2}\left(\ln\frac{x}{2\pi}-1\right)+\psi^{(-2)}(x)-\frac{\ln A^{12}x}{12} =  \frac{1}{720 x^2}-\frac{1}{5040 x^4} +\ldots 
\eea
In the regime of strong magnetic fields, where
the dynamical mass becomes very small compared to $\sqrt{2|q_rB_r|}$, i.e. 
$x\ll 1$, the momentum typically exchanged in scattering processes and hence the renormalization scale will be dominated by the scale set by magnetic field, 
not by the mass. Presumably, the physically most appropriate choice for the renormalization scale is thus a combination
of the mass and the magnetic field.

For our purpose, however, it is only important to notice that $\ell$ is a scale at which we
evaluate the final physical result after minimizing the free energy: when we take the derivative of the free
energy with respect to the dynamical mass $M$, we do so at fixed $\ell$; and, when we determine the onset of nuclear matter
we compare the free energy of the vacuum with the free energy of nuclear 
matter at the same value of $\ell$. Therefore, we do not have to specify the renormalization scale, and the terms independent of $M$, i.e., the first two 
terms in the second line of Eq.\ (\ref{Omrenorm}) play no role.
(When we discuss the $B$-independent contribution of the Dirac
sea in appendix $B$, we choose the vacuum nucleon mass as a 
renormalization scale, following the $B = 0$ literature.)

\section{Models}
\label{sec:models}

In this section, we introduce the two different models and write down the equations for minimizing the free energy, whose solution we discuss 
in Secs.\ \ref{sec:vacmasses} and \ref{sec:onset}. The models differ mainly in how they treat chiral symmetry breaking: in the Walecka model, the nucleons have a
given vacuum mass (in the absence of a magnetic field) which is a parameter of the model. Chiral symmetry is broken by construction, the model 
cannot describe chiral symmetry restoration. In the extended linear sigma model,
there is no mass term for the nucleons in the Lagrangian. The mass is generated dynamically by the formation of a chiral condensate. Therefore, the effect of magnetic 
catalysis can be seen very directly, by the $B$-dependence of the chiral condensate, whereas in the Walecka model it can only be seen in an indirect way, by the
$B$-dependence of the nucleon mass. 

The Lagrangian of both models has the form 
\be
{\cal L} = {\cal L}_N + {\cal L}_{\rm mes} + {\cal L}_I + {\cal L}_{\rm field}\, , 
\ee
where ${\cal L}_N$ describes free nucleons and their coupling to the magnetic field, 
${\cal L}_{\rm mes}$ the mesons and their (self-)interactions, ${\cal L}_I$ the (Yukawa-)interaction 
between the nucleons and mesons, and ${\cal L}_{\rm field}=-\frac{1}{4}F_{\mu\nu}F^{\mu\nu}$ is the free field part, giving rise to the $B^2$ term in the
free energy (\ref{Omgen}). 
The nucleonic part ${\cal L}_N$ includes the covariant derivative $D_\mu = \partial_\mu +i Q A_\mu$, where $A_\mu = (0,yB,0,0)$ accounts for a 
homogeneous background magnetic field in the $z$-direction, and $Q={\rm diag}(q_1,q_2)$ is the electric charge matrix in isospin space. 
For ordinary nuclear matter, $q_1=0$ (neutrons) and $q_2=e$ (protons).

\subsubsection{Walecka model}

In the Walecka model, the nucleonic part of the Lagrangian is
\be
{\cal L}_N = \bar{\psi}(i\gamma^\mu D_\mu -m_N +\gamma^0\mu)\psi \, , 
\ee
where $\psi$ is the nucleon spinor in isospin space, $m_N=939\,{\rm MeV}$ is the vacuum mass of the nucleons, and $\mu$ is the baryon chemical potential. 
The mesonic part contains 
the sigma and omega mesons, including scalar self-interactions,
\be \label{Lmeswal}
{\cal L}_{\rm mes} = \frac{1}{2}\left(\partial_\mu\sigma\partial^\mu\sigma - m_\sigma^2\sigma^2\right) - \frac{b}{3} m_N(g_\sigma \sigma)^3 
- \frac{c}{4}(g_\sigma\sigma)^4- \frac{1}{4}\omega_{\mu\nu}\omega^{\mu\nu} + \frac{1}{2}m_\omega^2 \omega_\mu\omega^\mu  \, , 
\ee
where $\omega_{\mu\nu}\equiv \partial_\mu\omega_\nu - \partial_\nu\omega_\mu$. The omega mass is $m_\omega = 782\, {\rm MeV}$ and for the sigma mass we 
use $m_\sigma = 550 \, {\rm MeV}$. The meson-nucleon interactions are given by
\be \label{LIwal}
{\cal L}_I = g_\sigma\bar{\psi}\sigma \psi - g_\omega \bar{\psi}\gamma^\mu\omega_\mu \psi \, .
\ee  
We employ the mean-field approximation, i.e., we neglect the fluctuations around the mesonic background fields $\bar{\sigma}$ and 
$\bar{\omega}_0$, which we assume to 
be uniform in space and time. One may ask whether a vector condensate $\bar{\omega}_i$ develops in the presence of a magnetic field. Minimizing the free energy 
with respect to this condensate shows that it must be proportional to a baryon current. Since there is no baryon current in an externally applied magnetic field
(only with a chiral imbalance can there be a current due to the chiral magnetic effect), the vector condensate vanishes. 

The dynamical nucleon mass and the effective chemical potential are 
\be
M_N = m_N - g_\sigma\bar{\sigma} \, , \qquad  \mu_* = \mu-g_\omega\bar{\omega}_0 \, .
\ee
In our approach with isospin symmetric bare masses and interactions, there is only one $M_N$, for both neutrons and protons. 
The isospin-breaking difference in electric charges leads to different excitation energies in a magnetic field, but not to different mass parameters $M_N$. 

The coupling constants of the model are fitted to reproduce the properties of nuclear matter at saturation in the absence of a magnetic field, 
namely the saturation density $n_0$, the binding energy $E_{\rm bind}$, the compression modulus $K$, and the dynamical mass at saturation,  
\be \label{nEKM}
n_0 = 0.153\,{\rm fm}^{-3} \, , \qquad E_{\rm bind} = -16.3 \, {\rm MeV} \, , \qquad K = 250\,{\rm MeV} \, , \qquad M_N = 0.8\,m_N \, ,
\ee
which leads to a chemical potential $\mu_0=922.7\,{\rm MeV}$ at saturation. The resulting values for the coupling constants are given in Table \ref{table0} 
in appendix \ref{app0}, where we also explain the fitting procedure for both models. 
There is some uncertainty especially in the compression modulus and the
effective mass and thus some arbitrariness in our choice of their values; 
the compression modulus is known to be in the range of $(200 - 300)\,{\rm MeV}$ \cite{Blaizot:1995zz,Youngblood:2004fe}, 
while the effective mass is in the range of $(0.7-0.8)\,m_N$ \cite{1989NuPhA.493..521G,Johnson:1987zza,Li:1992zza,glendenningbook}, 
possibly smaller \cite{Jaminon:1989wj,Furnstahl:1997tk}. We have chosen a value on the upper end of that range because in the extended linear 
sigma model lower values tend to be in conflict with vacuum properties, see remarks at the end of appendix \ref{app0}. 

The tree level potential $U$ is
\be \label{Uwalecka}
U = \frac{1}{2}m_\sigma^2\bar{\sigma}^2 + \frac{b}{3}m_N(g_\sigma\bar{\sigma})^3 + \frac{c}{4}(g_\sigma\bar{\sigma})^4 - \frac{1}{2} m_\omega^2\bar{\omega}_0^2 \, ,
\ee
and the equations we have to solve in order to minimize the free energy are
\be
\frac{\partial\Omega}{\partial \bar{\sigma}} =\frac{\partial\Omega}{\partial \bar{\omega}_0}=0 \, .
\ee
Using the expressions from Sec.\ \ref{sec:free} for the Dirac sea contribution and the matter contribution to the free energy, these equations can be written as
\begin{subequations} \label{statwal}
\bea
n_s &=& \frac{m_N-M_N}{g_\sigma^2/m_\sigma^2} + bm_N(g_\sigma \sigma)^2 + c(g_\sigma\sigma)^3 +\frac{|qB|M_N}{2\pi^2}\left[x(1-\ln x)
+\frac{1}{2}\ln\frac{x}{2\pi} + \ln\Gamma(x)\right]  \, , \label{statwal1} \\[2ex]
n &=& \frac{\mu-\mu_*}{g_\omega^2/m_\omega^2} \label{statwal2} \, .
\eea
\end{subequations}
The derivative of the sea contribution (\ref{Omrenorm}) was taken at fixed renormalization scale $\ell$, as argued below Eq.\ (\ref{Omrenorm}), and
we have introduced the scalar and baryon densities, 
\begin{subequations} \label{nBns}
\bea
n_s &=& \frac{\partial \Omega_{N,{\rm mat}}}{\partial M_N}
= \frac{|qB|}{2\pi}\sum_{e=\pm}\sum_{\nu=0}^\infty \alpha_\nu\int_{-\infty}^\infty \frac{dk_z}{2\pi} \frac{M_N}{\epsilon_{k,\nu}}f(\epsilon_{k,\nu}-e\mu_*) 
+ 2 \sum_{e=\pm}\int\frac{d^3{\bf k}}{(2\pi)^3} \frac{M_N}{\epsilon_{k}}f(\epsilon_{k}-e\mu_*)\non[2ex]
&\stackrel{T=0}{\longrightarrow}& \Theta(\mu_*-M_N)\left[\frac{|qB|M_N}{2\pi^2}\sum_{\nu=0}^{\nu_{\rm max}}\alpha_\nu \ln\frac{\mu_*+k_{F,\nu}}{\sqrt{M_N^2+2\nu|qB|}} +
\frac{M_N}{2\pi^2}\left(k_F\mu_*-M_N^2\ln\frac{k_F+\mu_*}{M_N}\right)\right]\, , \\[2ex]
n &=& -\frac{\partial \Omega_{N,{\rm mat}}}{\partial \mu} = 
\frac{|qB|}{2\pi}\sum_{e=\pm} e \sum_{\nu=0}^\infty \alpha_\nu \int_{-\infty}^\infty \frac{dk_z}{2\pi}f(\epsilon_{k,\nu}-e\mu_*)
+ 2 \sum_{e=\pm} e\int\frac{d^3{\bf k}}{(2\pi)^3} f(\epsilon_{k}-e\mu_*)\non
&\stackrel{T=0}{\longrightarrow}& \Theta(\mu_*-M_N)\left(\frac{|qB|}{2\pi^2}\sum_{\nu=0}^{\nu_{\rm max}}\alpha_\nu k_{F,\nu} + \frac{k_F^3}{3\pi^2}\right)  \, , 
\eea
\end{subequations}
where 
\be
f(x) = \frac{1}{e^{x/T}+1} 
\ee
is the Fermi distribution function. Scalar and baryon densities each contain contributions from protons and neutrons, 
while the $B$-dependent term in Eq.\ (\ref{statwal1}) originates only from the protons. Since in all relevant 
terms the renormalized magnetic field only appears in the combination $q_rB_r=q B$, we can choose the more compact notation $qB$, but keep in mind that the 
renormalization explained in Sec.\ \ref{sec:renorm} has been carried out.

\subsubsection{Extended linear sigma model}

In the extended linear sigma model, the nucleonic part of the Lagrangian is  
\be
{\cal L}_N = \bar{\Psi}(i\gamma^\mu D_\mu +\gamma^0\mu)\Psi \, , 
\ee
where 
\be
\Psi = \left(\begin{array}{c} \psi_1 \\ \psi_2 \end{array}\right) 
\ee
is a nucleon doublet where each of the components $\psi_1$ and $\psi_2$ (themselves being doublets in isospin space) transform oppositely under chiral 
transformations (``mirror assignment'') \cite{Detar:1988kn,Jido:1998av,Jido:2001nt,Zschiesche:2006zj,Gallas:2009qp,Gallas:2011qp,Heinz:2013hza}. 
The Dirac operator $i\gamma^\mu D_\mu +\gamma^0\mu$ is diagonal in this ``mirror space''.  

The mesonic part is \cite{Heinz:2013hza} (see also Refs.\ \cite{Gallas:2009qp,Parganlija:2010fz} for a more complete version)
\bea \label{LmeseLSM}
{\cal L}_{\rm mes} &=& \frac{1}{2}\left(\partial_\mu\sigma\partial^\mu\sigma + \partial_\mu\bm{\pi}\cdot\partial^\mu\bm{\pi}\right) 
+ \frac{1}{2}\partial_\mu\chi\partial^\mu\chi
+\frac{1}{2}m^2(\sigma^2+\pi^2)-\frac{\lambda}{4}(\sigma^2+\pi^2)^2+\epsilon\sigma - \frac{1}{2}m_\chi^2\chi^2 +g\chi(\sigma^2+\pi^2) \non[2ex]
&&- \frac{1}{4}\omega_{\mu\nu}\omega^{\mu\nu}+\frac{1}{2}m_\omega^2\omega_\mu\omega^\mu  \, . 
\eea
Here, $\sigma$ has a different meaning than the $\sigma$ in the Walecka model, where it is a massive mode with mass $m_\sigma$ in the chirally broken 
phase. The mass term for $\sigma$ in the present model has the ``wrong'' sign in order to model chiral symmetry breaking, i.e., $m$ in Eq.\ (\ref{LmeseLSM}) has a 
completely different meaning than $m_\sigma$ in Eq.\ (\ref{Lmeswal}). In the presence of a chiral condensate, the mass eigenstates of the scalars are 
linear combinations of the fields $\chi$ and $\sigma$. The lighter of these states can be identified with the $f_0(500)$ resonance \cite{Gallas:2011qp},
and thus plays the role of the $\sigma$ of the Walecka model; the heavier one is associated with $f_0(1370)$. This mixing is also discussed in appendix \ref{app0}.

The omega meson appears in both models in the same way, i.e., also here we have $m_\omega = 782\,{\rm MeV}$.
The meson-nucleon interactions are 
\be \label{LIeLSM}
{\cal L}_{\rm I} = \bar{\Psi}\left(\begin{array}{cc} -\displaystyle{\frac{\hat{g}_1}{2}}\sigma - g_\omega \gamma_\mu\omega^\mu & a\chi\gamma^5 \\[2ex] -a\chi\gamma^5 
& -\displaystyle{\frac{\hat{g}_2}{2}}\sigma - g_\omega \gamma_\mu\omega^\mu \end{array}\right)\Psi \, .
\ee
In the presence of a condensate $\bar{\chi}$, the off-diagonal components give rise to a chirally invariant mass 
term $-a\bar{\chi}(\bar{\psi}_2\gamma^5\psi_1 - \bar{\psi}_1\gamma^5\psi_2)$ with dynamically generated mass $a\bar{\chi}$. The remaining interactions in ${\cal L}_I$ 
are mediated by sigma and omega, where, for the latter, we have assumed the coupling constants for both components $\psi_1$ and $\psi_2$ to be the same, 
i.e., there is only a single $g_\omega$. 

This Lagrangian has more parameters than the Walecka model, leading to a more realistic description of vacuum properties of QCD \cite{Parganlija:2010fz,Janowski:2011gt}. 
Here we are interested in the
properties of nuclear matter, and for a sensible comparison between the two models in the presence of a magnetic field we need both models to reproduce the same 
saturation properties of nuclear matter in the absence of a magnetic field. Therefore, we fit the parameters of the extended linear sigma model 
also to the properties (\ref{nEKM}); this is explained in detail 
in appendix \ref{app0}, where also the numerical values for the parameters are given, see Table \ref{table0}.

With the help of the present model, it has been argued that the chiral condensate in nuclear matter
can become anisotropic (``chiral density wave''). This anisotropic state of nuclear matter has been discussed in Ref.\ \cite{Heinz:2013hza} in the absence
of a magnetic field with the ansatz $\bar{\sigma} = \phi\cos(2fx)$, $\bar{\pi}_3=\phi\sin(2fx)$, with $\phi$ and $f$ to be determined dynamically. 
It is an interesting question whether a chiral density wave (or a more complicated, inhomogeneous 
structure) also occurs in the presence of a magnetic field. 
Since the magnetic field already breaks rotational symmetry, one might expect the chiral density wave to be even more favored in this case. This expectation is supported 
by model calculations where quarks, not nucleons, are the degrees of freedom \cite{Rebhan:2008ur,Frolov:2010wn,Preis:2010cq}. Here we will ignore  the 
possibility of a chiral density wave for simplicity, leaving a study with background magnetic field {\it and} chiral density wave as a next step for the future. 
We thus proceed with a uniform, rotationally symmetric ansatz for 
the chiral condensate $\bar{\sigma}$, and set $\bar{\pi}_i=0$. As in the Walecka model, we work in the mean-field approximation, neglecting all fluctuations around 
the condensates $\bar{\sigma}$ and $\bar{\omega}_0$.  

In the absence of a magnetic field, the determinant of the inverse fermionic propagator in momentum space $G^{-1}$ is
\be
{\rm det}\,G^{-1} = \Big\{ (a\bar{\chi})^4 -2(a\bar{\chi})^2\left[(k_0+\mu_*)^2-\left(k^2+m_1m_2\right)\right]
+\left[(k_0+\mu_*)^2-\left(k^2+m_1^2\right)][(k_0+\mu_*)^2-\left(k^2+m_2^2\right)\right]\Big\}^2 \, ,
\ee
where $\mu_* = \mu-g_\omega\bar{\omega}_0$, as in the Walecka model, and we have abbreviated $m_1\equiv \hat{g}_1\bar{\sigma}/2$, $m_2\equiv \hat{g}_2\bar{\sigma}/2$. 
The zeros of the determinant are $\epsilon_{k,i}-\mu_*$ with the excitation energies $\epsilon_{k,i} = \sqrt{k^2+M_i^2}$ ($i=N,N^*$), where 
\be \label{MNMNs}
M_{N,N^*}= \pm \frac{\hat{g}_1-\hat{g}_2}{4}\bar{\sigma} + \sqrt{(a\bar{\chi})^2+\left(\frac{\hat{g}_1+\hat{g}_2}{4}\right)^2\bar{\sigma}^2}   \, .
\ee
The degeneracy between the masses of the nucleon $M_N$ and its chiral partner $M_{N^*}$ -- identified with the resonance $N(1535)$ -- 
is broken by the chiral condensate $\bar{\sigma}$. Since we shall be interested in the zero-temperature onset of nuclear matter, which 
occurs at energies well below $M_{N^*}$, the nucleonic states of the chiral partner will not be occupied in any of our results. We shall see, however, that they 
play a non-negligible role in the sea contribution for large magnetic fields.   

Including a magnetic field is straightforward since it couples to both components $\psi_1$ and $\psi_2$ equally. Therefore, the excitations for the charged nucleons assume the form (\ref{epskn}), with $M$ replaced by $M_N$ and $M_{N^*}$.

The tree-level potential is
\be \label{UeLSM}
U  = -\frac{1}{2}m^2\bar{\sigma}^2 - \epsilon\bar{\sigma} + \frac{\lambda}{4}\bar{\sigma}^4  -\frac{1}{2}m_\omega^2\bar{\omega}_0^2 
+\frac{1}{2}m_\chi^2\bar{\chi}^2- g\bar{\chi}\bar{\sigma}^2 \, ,
\ee
and the three condensates are determined from  
\be
\frac{\partial\Omega}{\partial \bar{\sigma}}  = \frac{\partial\Omega}{\partial \bar{\chi}}=\frac{\partial\Omega}{\partial \bar{\omega}_0}= 0 \, .
\ee
In the presence of a magnetic field, these equations become
\begin{subequations} \label{stateLSM}
\bea
\epsilon+m^2\bar{\sigma}-\lambda\bar{\sigma}^3+2g\bar{\chi}\bar{\sigma}&=& \sum_{i=N,N^*}
\left\{-\frac{M_i|qB|}{2\pi^2}\left[x_i(1-\ln x_i)+\frac{1}{2}\ln\frac{x_i}{2\pi}+\ln\Gamma(x_i)
\right]+\frac{\partial\Omega_{N,{\rm mat}}}{\partial M_i}\right\}\frac{\partial M_i}{\partial\bar{\sigma}} \, , \label{stateLSM1}\\[2ex]
g\bar{\sigma}^2-m_\chi^2\bar{\chi}&=& \sum_{i=N,N^*}\left\{-\frac{M_i|qB|}{2\pi^2}\left[x_i(1-\ln x_i)+\frac{1}{2}\ln\frac{x_i}{2\pi}+\ln\Gamma(x_i)
\right]+\frac{\partial\Omega_{N,{\rm mat}}}{\partial M_i}\right\}\frac{\partial M_i}{\partial\bar{\chi}} \, , \label{stateLSM2}\\[2ex]
n &=& \frac{\mu-\mu_*}{g_\omega^2/m_\omega^2} \, .\label{stateLSM3}
\eea
\end{subequations}
Here, the matter part of the free energy $\Omega_{N,{\rm mat}}$ has contributions from both protons and neutrons, as given in Eqs.\ (\ref{OmNmat}) 
and (\ref{OmNmat0}), each generalized to include both nucleon states $i=N,N^*$. The total baryon density $n$ is then defined as usual by the 
(negative of the) derivative of $\Omega_{N,{\rm mat}}$ with respect to $\mu$. The $B$-dependent sea contribution in the 
curly brackets of Eqs.\ (\ref{stateLSM1}) and (\ref{stateLSM2}) originates from the proton and its chiral partner, with 
$x_i=M_i^2/(2|qB|)$ being the obvious generalization of the abbreviation introduced in Eq.\ (\ref{x}). Expanding for large $x_i$ yields
\be\label{largeM}
x_i(1-\ln x_i)+\frac{1}{2}\ln\frac{x_i}{2\pi}+\ln\Gamma(x_i) = \frac{1}{12 x_i} - \frac{1}{360 x_i^3} + {\cal O}\left(\frac{1}{x_i^5}\right) \, .
\ee
Therefore, for a given magnetic field $|qB|$, the contribution of heavy states with masses $M^2 \gg |qB|$ is suppressed, as one might have expected.

\section{Results I: vacuum masses}
\label{sec:vacmasses}

\begin{figure}[t] 
\begin{center}
\hbox{\includegraphics[width=0.5\textwidth]{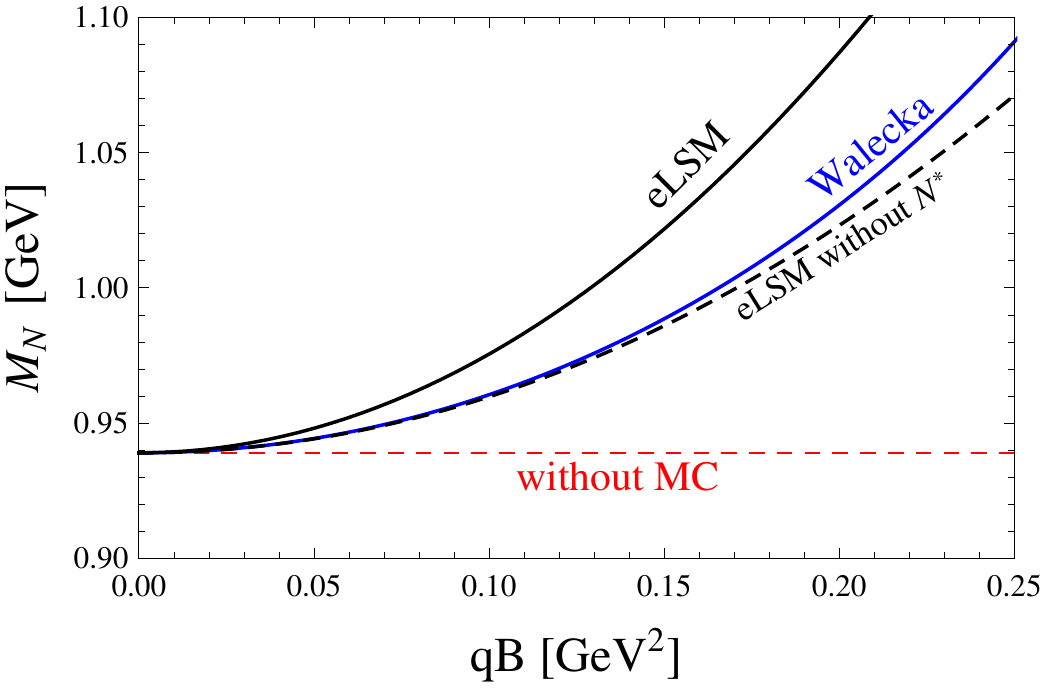}\includegraphics[width=0.5\textwidth]{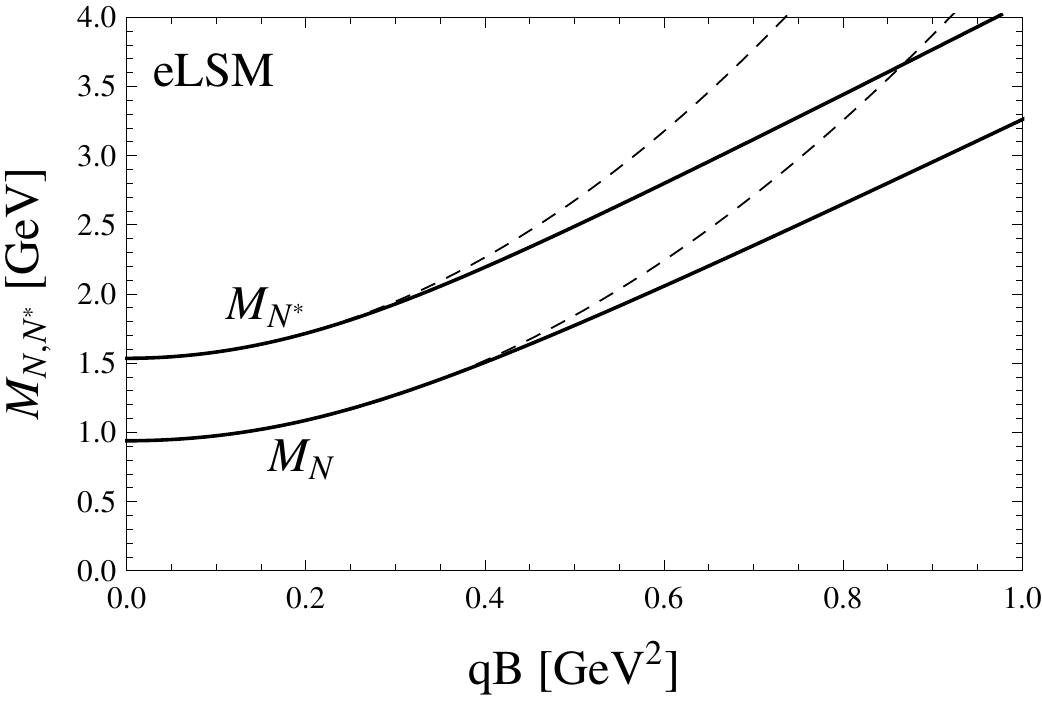}}
\caption{Left panel: dependence of the vacuum nucleon mass on the magnetic field in the Walecka model and the extended linear sigma model (eLSM). 
The increase of the vacuum mass is expected from magnetic catalysis (MC); neglecting the contribution of the Dirac sea that is responsible for magnetic catalysis 
leads to an incorrect constant vacuum mass (horizontal thin dashed line). The thick dashed line is the result of the extended linear sigma model 
without the sea contribution of the chiral partner of the nucleon $N^*$, showing that the main difference between the two models originates from that
contribution. 
Right panel: vacuum mass of the nucleon $M_N$ and its chiral partner $M_{N^*}$ according to the extended linear sigma model on a larger scale for the magnetic field,
showing a linear behavior for very strong fields. The dashed lines are the quadratic approximations for small fields. 
(If $q = e \simeq 0.30$, then $qB=0.1\,{\rm GeV}^2$ in natural Heaviside-Lorentz units corresponds to $B= 1.7\times 10^{19}\,{\rm G}$ in Gaussian units.)
}
\label{fig:vacmass}
\end{center}
\end{figure}

In this section we discuss the vacuum $\mu=T=0$ in both models, starting with the Walecka model. The Walecka model is a phenomenological model for nuclear matter, 
therefore we should not expect it to yield profound results for the QCD vacuum. Nevertheless, we shall see that it can account for magnetic catalysis. 
For $\mu=T=0$ we have  $n=n_s=0$, and thus Eq.\ (\ref{statwal2}) is trivially solved by $\bar{\omega}_0=0$. The remaining Eq.\ (\ref{statwal1}) has to be solved 
for $\bar{\sigma}$. For vanishing magnetic field there is one unique solution $\bar{\sigma}=0$ for our parameter choice of $m_\sigma$, $b$, and $c$. 
The uniqueness of the solution can easily get lost for slightly different (still physically sensible) parameters. For instance, fitting the parameters to an 
effective nucleon mass at saturation $M_N=0.78\, m_N$ instead of $M_N=0.8\, m_N$ with all other properties in Eq.\ (\ref{nEKM}) kept fixed, 
leads to a potential that allows for three solutions. The physical solution $\bar{\sigma}=0$ is then a local minimum, and the global minimum at some large negative value
of $\bar{\sigma}$ has to be ignored. 

For nonzero magnetic field, $\bar{\sigma}$ assumes negative values, leading to an increased nucleon mass. For small magnetic fields, this increase is quadratic,
\bea \label{smallBwal}
\frac{M_N(\mu=T=0)}{m_N} \simeq 1+ \frac{g_\sigma^2|qB|^2}{12\pi^2m_N^2m_\sigma^2} 
\simeq 1+ \left(\frac{|qB|}{0.67\,{\rm GeV}^2}\right)^2  \, .
\eea
The numerical solution is shown in the left panel of Fig.\ \ref{fig:vacmass}.
For very large magnetic fields, the model in its present form cannot be trusted. This is most obvious with the slightly different parameter set 
just mentioned, where $M_N=0.78\, m_N$: in that case, the physical solution ceases to exist at around $qB\simeq 0.3 \,{\rm GeV}^2$. 
This artifact might be cured by $B$-dependent meson masses. In the mean-field approach, 
neither of the two models we use predicts any effect of the magnetic field on the meson masses. It could be computed from loop corrections, or 
from a more microscopic approach. Here, throughout the paper, we shall neglect such an effect on the meson masses and the coupling constants. 

Also in the extended linear sigma model, we obviously have $\bar{\omega}_0=0$ in the vacuum. For $\mu=T=B=0$ the right-hand sides of Eqs.\ (\ref{stateLSM1}) 
and (\ref{stateLSM2})
vanish. As a consequence, Eq.\ (\ref{stateLSM2}) yields $\bar{\chi}$ as a simple function of $\bar{\sigma}$ and Eq.\ (\ref{stateLSM1}) becomes a 
cubic equation for $\bar{\sigma}$ with three solutions. The global minimum of the free energy [which in this case is simply the tree-level potential (\ref{UeLSM})] is
\be \label{barsigma0}
\bar{\sigma}(\mu=T=B=0) = \frac{2m}{\sqrt{3}\sqrt{\lambda-\frac{2g^2}{m_\chi^2}}}
\cos\left[\frac{1}{3}\arccos\left(\frac{3\sqrt{3}\epsilon}{2m^3}\sqrt{\lambda-\frac{2g^2}{m_\chi^2}}\right)\right]  \simeq 154.3 \, {\rm MeV} \, ,
\ee
and $\bar{\chi}=g\bar{\sigma}^2/m_\chi^2\simeq 27.9\, {\rm MeV}$. By construction, this yields the two nucleon masses $m_N \equiv M_N(\mu=T=B=0) =  939\,{\rm MeV}$
and $m_{N^*} \equiv M_{N^*}(\mu=T=B=0) = 1535 \, {\rm MeV}$. The evaluation of Eqs.\ (\ref{stateLSM1}) and (\ref{stateLSM2}) shows that, for $\mu=T=0$, the 
chiral condensate $\bar{\sigma}$ increases quadratically with $qB$. This is in agreement with chiral perturbation theory \cite{Agasian:2001hv}
(in the chiral limit, the behavior is linear in the magnetic field \cite{Shushpanov:1997sf,Cohen:2007bt}), the quark-meson model \cite{Gatto:2012sp}, 
and the holographic Sakai-Sugimoto model \cite{Callebaut:2013ria}; see Ref.\ \cite{Bali:2012zg} for a comparison of lattice QCD results with the
various model predictions. 

We find that also $\bar{\chi}$ and the nucleon masses increase quadratically with $qB$. One can determine the coefficients for the quadratic terms 
analytically. They are very complicated and not very instructive combinations of the parameters of the model. Therefore, we simply give the numerical
result for the nucleon mass, 
\be \label{smallBeLSM}
\frac{M_N(\mu=T=0)}{m_N} \simeq  1+\left(\frac{|qB|}{0.51\, {\rm GeV}^2}\right)^2 \, .
\ee
The full numerical result is shown in Fig.\ \ref{fig:vacmass}. In the right panel, we show the numerical result for both nucleon masses on a large scale of 
the magnetic field. We observe a linear increase of the vacuum masses for very strong magnetic fields. In the left panel, the results of the two models are compared. 
Interestingly, if we ignore the chiral partner
(i.e., remove its contribution from the sea terms by hand while keeping all parameters fixed) the two models are in much better agreement, suggesting that the 
main difference between the models comes from the additional hadronic state in the Dirac sea. We have seen in Eq.\ (\ref{largeM}) that heavy states with mass $M$ do not 
contribute if $M^2\gg |qB|$. It turns out, however, that the magnetic fields considered here are sufficiently large for the sea contribution of both 
the nucleon and its chiral partner to have a sizable effect on the vacuum masses. 
It can be expected that in a more complete treatment, 
including other charged hadronic states such as pions or hyperons, our results will further be changed quantitatively.

\begin{figure}[t] 
\begin{center}
\hbox{\includegraphics[width=0.5\textwidth]{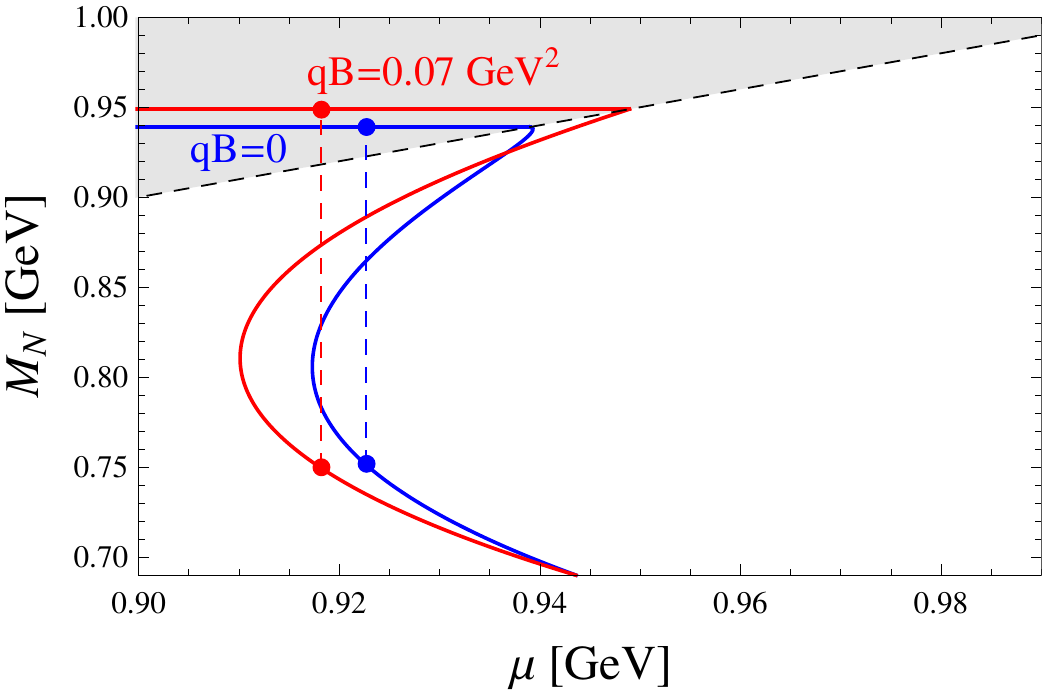}\includegraphics[width=0.5\textwidth]{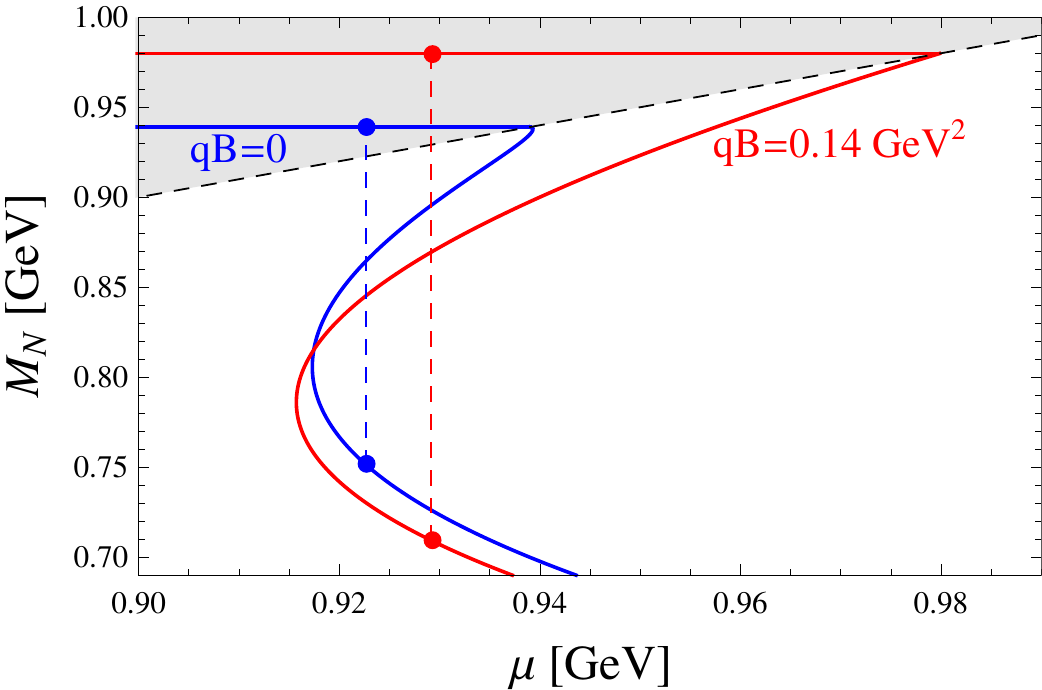}}
\caption{Zero-temperature solutions for the nucleon mass in the Walecka model in the vicinity of the onset of nuclear matter 
for two different magnetic fields, compared to the solution in the absence of a magnetic field.
The vertical dashed lines indicate the onset of nuclear matter, where the effective mass decreases discontinuously. 
The vacuum mass (horizontal segments of the curves) is increased for 
both magnetic fields due to magnetic catalysis. With respect to the onset, the two panels show two different cases: in the left (right) panel, the 
critical chemical potential for the onset is smaller (larger) than without magnetic field. The (black) dashed line is defined by $\mu=M_N$, such that the shaded 
area corresponds to the vacuum. In the extended linear sigma model, the results look qualitatively always like in the right panel.}
\label{fig:multi}
\end{center}
\end{figure}

\section{Results II: nuclear matter onset}
\label{sec:onset}

We now include the medium terms and discuss the onset of nuclear matter at zero temperature as a function of the magnetic field in both models. 
We start by numerically solving Eqs.\ (\ref{statwal}) for $\bar{\sigma}$, $\bar{\omega}_0$ (Walecka model) and 
Eqs.\ (\ref{stateLSM}) for $\bar{\sigma}$, 
$\bar{\omega}_0$, $\bar{\chi}$ (extended linear sigma model) for a fixed magnetic field. 

In Fig.\ \ref{fig:multi} we show the solutions for the nucleon mass in the vicinity of the nuclear matter onset for the Walecka model. 
This plot is helpful for an 
understanding of the structure of the solutions: at $T=0$, there is obviously no contribution from the medium if the effective chemical potential $\mu_*$ 
is smaller than the nucleon mass. Now, since $\bar{\omega}_0= 0$ and thus $\mu_*=\mu$ in the vacuum, the same is true for the chemical potential 
$\mu$. Therefore, in the shaded area only the vacuum solution exists. This solution does not depend on $\mu$ and is thus given by a horizontal 
line. The vacuum solution increases monotonically with the magnetic field, i.e., for any nonzero magnetic field the horizontal line lies above the $qB=0$ line.
This is magnetic catalysis. 

In the unshaded area, 
the medium terms contribute and $\bar{\omega}_0\neq 0$. There is a regime where three solutions and thus three values for the nucleon mass exist. 
In this regime a first-order phase transition occurs, the onset of nuclear matter, as indicated by the vertical dashed lines. 
At the onset, the free energy of nuclear matter starts to become smaller than the free energy of the vacuum. Therefore, the onset is determined 
by requiring the free energies of the vacuum and nuclear matter to be equal. 

\begin{figure}[t] 
\begin{center}
\hbox{\includegraphics[width=0.5\textwidth]{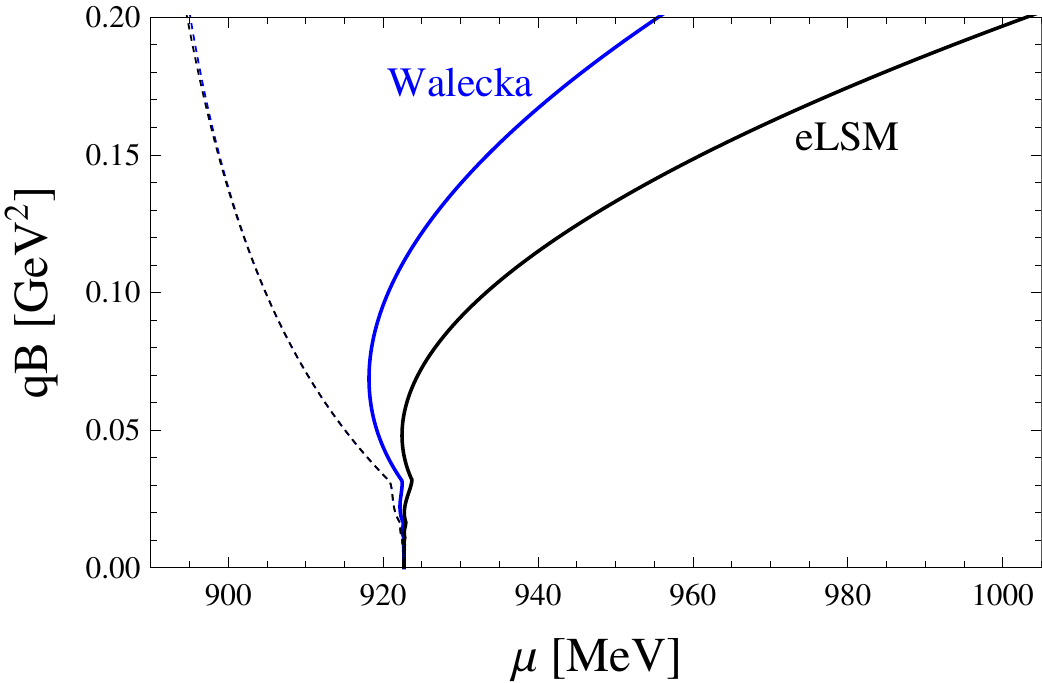}\includegraphics[width=0.5\textwidth]{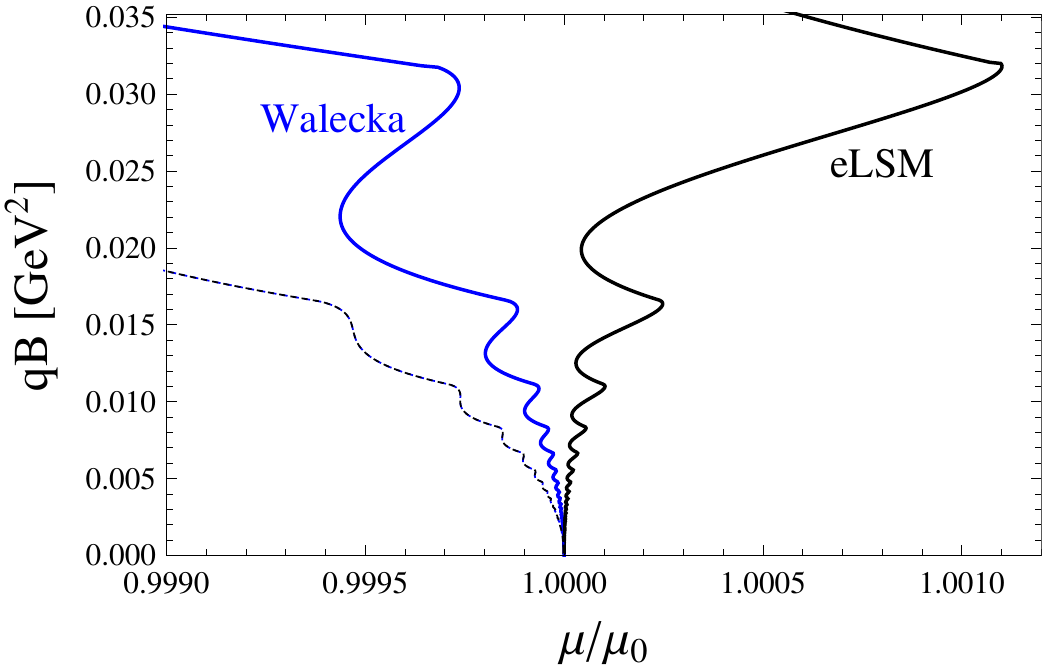}}
\caption{Onset for nuclear matter in the presence of a background magnetic field. The thick lines show our results for the Walecka model and the 
extended linear sigma model (eLSM). For comparison, the two thin dashed lines (barely distinguishable from each other) show the result within the same models, but without 
magnetic catalysis, i.e., ignoring the $B$-dependent sea contribution, as done in the previous literature. The right panel is a zoom-in to small 
magnetic fields and shows the oscillations due to the Landau levels (here, $\mu_0\simeq 922.7 \,{\rm MeV}$ is the 
chemical potential for the onset in the absence of a magnetic field).}
\label{fig:onset}
\end{center}
\end{figure}

The figure shows two qualitatively different cases: in the left panel, the onset occurs ``earlier''
than for vanishing magnetic field, even though the vacuum mass is enhanced; at the magnetic field chosen here, this ``inverse'' effect is most pronounced 
(see also Fig.\ \ref{fig:onset}). In the right panel the onset occurs ``later'' than in the absence of a magnetic field. 
Such a non-monotonic behavior is possible because the magnetic field 
also affects the binding energy of nuclear matter. The binding energy per nucleon (at saturation) is defined as the difference between the 
chemical potential -- the energy needed to put a single nucleon into the magnetized medium -- and the 
vacuum mass -- the energy needed to put the nucleon into the magnetized vacuum, 
\be
E_{\rm bind}(qB) = \mu_0(qB) - M_N(\mu=T=0,qB) \, , 
\ee
with $E_{\rm bind}(qB=0) \simeq -16.3\,{\rm MeV}$.
In Fig.\ \ref{fig:multi} the binding energy can be easily read off: by definition, it is the length of the horizontal segment of the curve between the point 
indicating the onset and the end of the shaded area.

In the linear sigma model, the situation of the left panel does not occur for our choice of parameters. The 
chemical potential for the onset of nuclear matter in a background magnetic field is 
always larger than that without a magnetic field, and the solution for the nucleon mass looks qualitatively the same as in the right panel.
In the vicinity of the onset, there is no population of the chiral partner because $\mu_*<M_{N^*}$,
and thus its medium contribution vanishes. 
In this model, there is another first-order phase transition at a larger value of the chemical potential 
where chiral symmetry is (approximately) restored. In this paper, we concentrate on the onset of nuclear matter, and leave the discussion of the chiral phase 
transition in the presence of a magnetic field for future studies.

\begin{figure}[t] 
\begin{center}
\hbox{\includegraphics[width=0.5\textwidth]{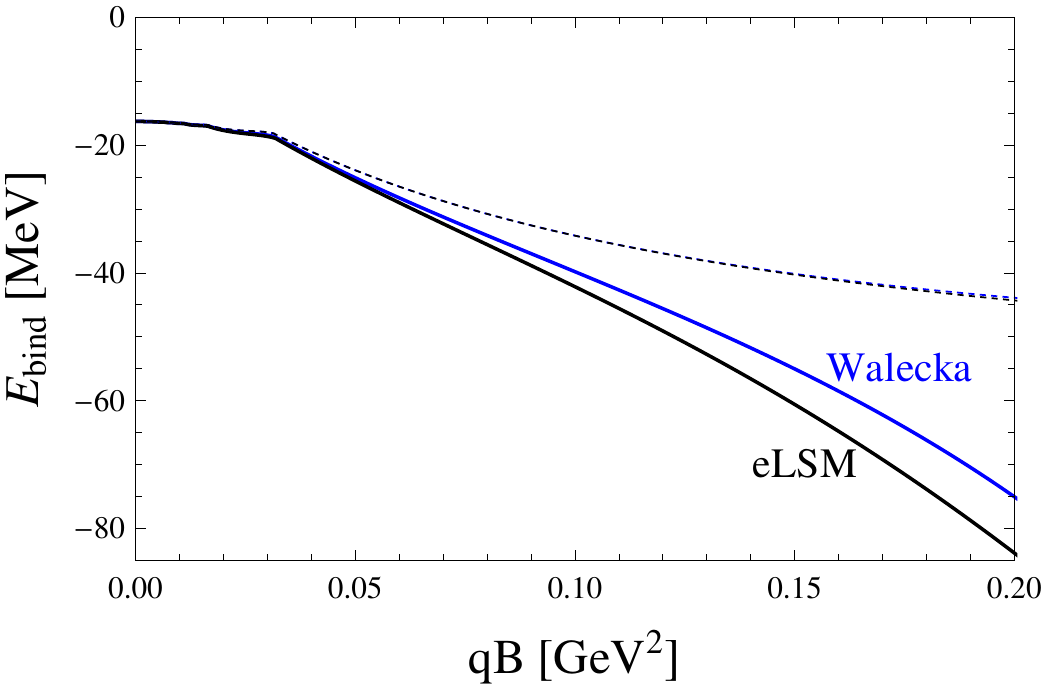}\includegraphics[width=0.5\textwidth]{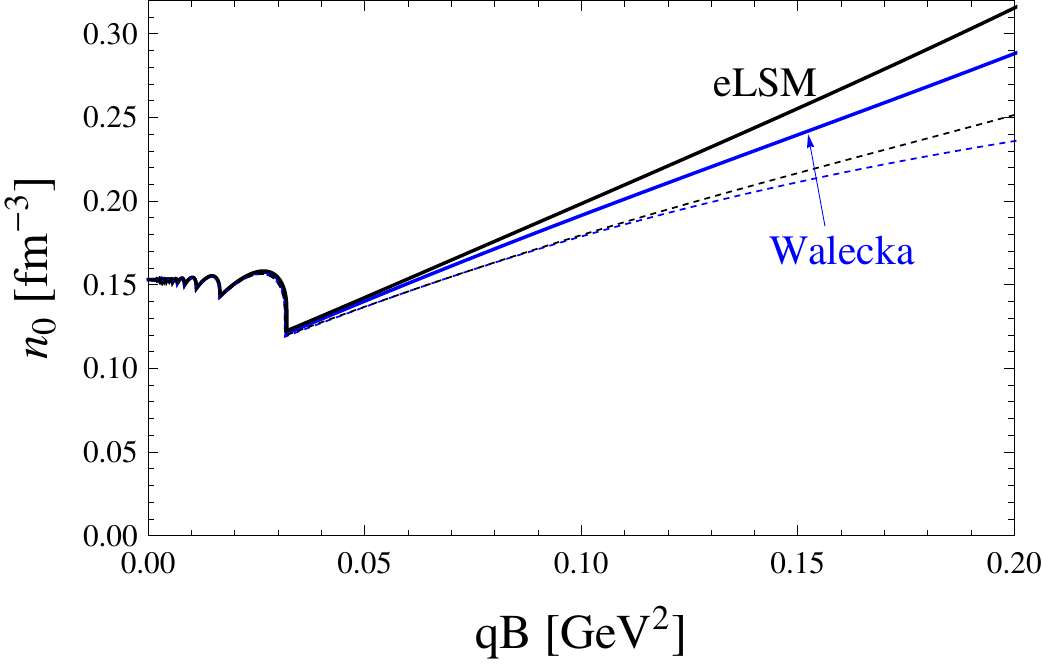}}
\caption{Binding energy (left) and baryon density (right) along the onset of nuclear matter for both models (solid lines). Again, the thin dashed lines
show the incorrect result with a $B$-independent vacuum mass of the nucleons. If rotated by $90^\circ$ 
and shifted by $m_N$, the dashed curves for the binding energy 
are identical to the dashed onset curves 
in Fig.\ \ref{fig:onset} because the entire effect of the magnetic field is given by the binding energy.}
\label{fig:nEb}
\end{center}
\end{figure}

We plot the critical chemical potential for all magnetic fields $qB<0.2\,{\rm GeV}^2$ in the left panel of Fig.\ \ref{fig:onset}. The right panel is a zoom-in to 
smaller magnetic fields. The corresponding binding energy and baryon density along the onset are shown in Fig.\ \ref{fig:nEb}.  We discuss the main 
observations separately.
\begin{itemize}

\item For small magnetic fields, the onset curve shows an oscillatory behavior. These oscillations are barely visible in the left panel
since they occur on a very small scale of the chemical potential. Their origin is the Landau level structure, i.e., depending on the strength of the magnetic 
field, saturated nuclear matter occupies different numbers of Landau levels, the larger the field the fewer the occupied levels. For all fields larger than about 
$qB\simeq 0.032\, {\rm GeV}^2$, saturated nuclear matter only occupies the lowest Landau level in both models. 

\item At sufficiently large magnetic fields, magnetic catalysis dominates the onset, nuclear matter becomes increasingly more
difficult to create because of the heavier nucleon mass. For comparison, we have plotted the incorrect result that is obtained without the Dirac sea contribution, i.e., 
with a constant vacuum mass (as indicated by the constant line in the left panel of Fig.\ \ref{fig:vacmass}). There is an obvious qualitative difference between these
results, with a difference of up to about $\sim 10\%$ for the onset chemical potential, $\sim 90\%$ for the binding energy, and $\sim 25\%$ for the saturation 
density at the largest magnetic field considered here, $qB = 0.2 \,{\rm GeV}^2$ (corresponding to $B\simeq 3.4\times 10^{19}\,{\rm G}$ for $q=e$).

It is instructive to compare this with the holographic Sakai-Sugimoto model, which, in a certain limit, is dual to large-$N_c$ QCD. In this model, 
the attractive nucleon-nucleon force and thus the binding energy is absent \cite{Kaplunovsky:2010eh}. 
As a consequence, the  onset is solely determined by the nucleon mass 
and therefore the critical chemical potential increases monotonically \cite{Bergman:2008qv,Preis:2011sp}. In Ref.\ \cite{Preis:2011sp}, the result of the Walecka 
model, however without taking into account the sea contribution, had been compared to the holographic result. It is no surprise that the present results, 
where the $B$-dependence of the vacuum mass is taken into account, are in better agreement to the Sakai-Sugimoto model since now both models account for 
magnetic catalysis. Nevertheless, due to the large-$N_c$ limit of the holographic calculation and the absence of an anomalous magnetic moment in the present work, 
a complete agreement should not be expected.

\item There is a significant difference between the two models, the linear sigma model having a larger onset chemical potential for all magnetic fields. 
This is in accordance with the 
observation made in Sec.\ \ref{sec:vacmasses}, where a stronger magnetic catalysis in the linear sigma model was pointed out and attributed to the presence of the
chiral partner of the nucleon. We have checked that the curves for the onset chemical potential of both models are almost exactly identical if we remove by hand the 
sea contribution of the chiral partner in the linear sigma model. 
(The models also coincide if the crucial sea terms are completely removed, as the dashed lines in the figures show.)

\item As already seen in Fig.\ \ref{fig:multi}, in the Walecka model there is a regime where the critical chemical potential is {\it lower} than in the absence of 
a magnetic field. This indicates that the binding energy is increased and can dominate the effect of the increasing vacuum mass, i.e., a magnetic field
can in principle also facilitate the creation of nuclear matter. Interestingly, the shape of the onset curve looks very similar to the chiral phase transition 
line obtained from a Nambu-Jona-Lasinio model or the Sakai-Sugimoto model \cite{Preis:2012fh}, showing  "inverse magnetic catalysis" \cite{Preis:2010cq}. 
In both cases, nuclear matter onset and chiral phase transition, the simple monotonic behavior at large magnetic fields becomes more complicated for 
smaller magnetic fields. 

\end{itemize}

\section{Conclusions}
\label{sec:summary}

We have discussed the onset of nuclear matter at zero temperature in the presence of a background magnetic field. Our main goal has been to investigate  the 
influence of the $B$-dependent contribution of the Dirac sea which had been omitted in previous studies about magnetized nuclear matter, but taken 
into account routinely
in very similar studies about quark matter. The physical meaning of this contribution can be interpreted as magnetic catalysis, an enhancement of the 
chiral condensate under the influence of a magnetic field. We have employed two different relativistic field-theoretical models in the mean-field approximation:
the Walecka model, where magnetic catalysis is seen
indirectly through an increased vacuum mass of the nucleon, and an extended linear sigma model, where the increase of the chiral condensate can be observed 
directly. 

If one ignores the effect of magnetic catalysis, creating nuclear matter becomes energetically less costly in a magnetic field, indicating 
an increased binding energy. But, if magnetic catalysis is properly taken into account, its effect dominates at large magnetic field 
and creating nuclear matter becomes energetically {\it more} costly, even though the binding energy is further increased. 
While this qualitative behavior is seen in both models, they differ quantitatively. 
We have shown that the main reason for the difference between the models can be attributed to the presence of the chiral partner of the nucleon in the 
extended linear sigma model. 
Its presence in the Dirac sea (it is too heavy to play a role in the medium part) leads to a stronger magnetic catalysis.

Our calculation isolates the effect of magnetic catalysis, but it has to be kept in mind that there are more effects that we have omitted. 
In view of expectations from full QCD, our results can be interpreted as follows. We know from the lattice that the chiral condensate increases monotonically 
with the magnetic field at zero temperature. This induces an increase in quark masses. Whether this leads to an increase in the vacuum masses of the nucleons
is not obvious because the interactions between the quarks will be modified by a strong magnetic field too. Our models do not know about the inner structure of the 
nucleons, and they show a simple monotonic increase of the vacuum masses. Modifications of the quark interactions may induce competing effects, seeking to reduce
the vacuum mass of the nucleons \cite{Andreichikov:2013pga}. (To our knowledge, vacuum masses of nucleons in a magnetic field have not yet been computed on the lattice.) 
In the models used here such effects can be included in an effective way through the anomalous magnetic moment of the nucleons, 
which indeed appears to counteract the effect of magnetic catalysis \cite{Preis:2011sp}.
In the present study, the resulting heavier nucleons suggest a larger critical chemical potential for the transition from vacuum to nuclear matter. 
But again, this is not the only important effect.
Now it is the interaction between the {\it nucleons} that is also modified by the magnetic field, and this effect {\it is} included in our models: 
we observe an increased binding energy, leading to a nontrivial behavior of the onset curve. 

Our study opens up various interesting questions that should be addressed in the future. Having just explained the main deficiency of our approach, it is clear that
in future studies the Dirac sea contribution {\it and} the anomalous magnetic moment should be taken into account. For the latter one should preferably go  
beyond the usually employed effective, non-renormalizable approach in order to allow for the renormalization of the sea contribution emphasized here. 
Moreover, our observation of the effect of the chiral partner shows that for quantitatively reliable predictions more charged hadronic states such as pions and
rho mesons, and possibly hyperons, need to be taken into account, even if some of these states are too heavy to be populated. It is also the scale set by the magnetic 
field, not only the chemical potential, to which their mass has to be compared in order to estimate their 
importance. 

For applications to compact stars, our results have to be extended to higher densities, and the conditions of beta equilibrium and charge neutrality 
have to be taken into account. This is more or less straightforward, and basically amounts to extending existing studies by including the $B$-dependent sea contribution,
possibly after generalizing our renormalized vacuum to the case of nonvanishing anomalous magnetic moments. It remains to be seen whether
magnetic catalysis has a sizable effect for instance on the equation of state and thus mass and radius of a compact star.  

It would also be interesting to consider  larger values of the baryon chemical potential within the extended linear sigma model. In contrast to the Walecka model,
the extended linear sigma model incorporates nuclear matter {\it and} the possibility of (approximate) restoration of chiral symmetry. One can thus use it to 
study the chiral phase transition in the presence of nuclear matter and a background magnetic field. This could be done, as a first step, in the present setup or,
in a more complicated scenario, after including an anisotropic chiral condensate in the form of a chiral 
density wave. One of the questions is whether there exists nuclear matter at very large magnetic field or whether the mesonic phase is directly
superseded by the chirally symmetric phase, as suggested by results within a holographic approach \cite{Preis:2011sp}.

\begin{acknowledgments}
We thank J.O.\ Andersen, G.\ Endr\H{o}di, M.\ Huang, E.\ Fraga, B.\ Kerbikov,  D.\ Parganlija, A.\ Rebhan, D.\ Rischke, I.\ Shovkovy, and M.\ Stephanov 
for useful comments and discussions.
This work has been supported by the Austrian science foundation FWF under project no.~P23536-N16, START project no.~Y435-N16, 
and by the NewCompStar network, COST Action MP1304.
\end{acknowledgments}

\appendix

\section{Parameters in the two models}
\label{app0}

In this appendix, we explain how we fit the parameters in the two models. 

In the Walecka model, we fix the meson masses to $m_\sigma=550\,{\rm MeV}$ and $m_\omega=782\,{\rm MeV}$, and the nucleon mass to $m_N=939\,{\rm MeV}$. The
remaining parameters are fitted to reproduce the properties of nuclear matter at saturation in the absence of a magnetic field, see Eq.\ (\ref{nEKM}). 
This is done as follows. First we note that the determination of $g_\omega$ decouples from the other parameters:
with $n=2k_{F}^3/(3\pi^2)$ and $\mu_0 = m_N+E_{\rm bind}$ we know the Fermi momentum $k_{F}$ and the chemical potential $\mu_0$ at saturation
because $n_0$ and $E_{\rm bind}$ are given. This allows us to compute 
$g_\omega\omega$ from $\mu-g_\omega\omega=\sqrt{k_{F}^2+M_N^2}$ since $M_{N}$ at saturation is given. 
Inserting the result into the $B\to 0$ limit of Eq.\ (\ref{statwal2}), we can determine $g_\omega$. 
For the determination of $g_\sigma$, $b$, and $c$, we solve the following three coupled equations: the remaining 
equation to minimize the free energy, i.e., Eq.\ (\ref{statwal1}), the condition that the pressure of nuclear matter vanishes at the onset, 
and the condition that the compression modulus assumes the value given in Eq.\ (\ref{nEKM}). For the compression
modulus at saturation we have \cite{glendenningbook}
\be \label{comp}
K = 9n\frac{\partial^2\epsilon}{\partial n^2} = \frac{6k_{F}^3}{\pi^2} \left(\frac{g_\omega}{m_\omega}\right)^2 + \frac{3k_{F}^2}{\mu_*}-\frac{6k_{F}^3}{\pi^2}
\left(\frac{M_{N}}{\mu_*}\right)^2\left[\frac{\partial^2 U}{\partial M_N^2}+\frac{2}{\pi^2}\int_0^{k_{F}}dk\,\frac{k^4}{\epsilon_k^3}\right]^{-1} \, ,
\ee
where $\epsilon$ is the energy density. The resulting parameters are given in Table \ref{table0}.

The extended linear sigma model has more parameters that have to be fitted. 
The mesonic part of the Lagrangian (\ref{LmeseLSM}) contains 6 parameters, $m$, $\epsilon$, $\lambda$, $m_\chi$, $g$, $m_\omega$. The omega
mass is again fixed to $m_\omega = 782\,{\rm MeV}$. To fix the other parameters, we compute the physical meson masses at tree level: we introduce 
condensates and fluctuations, $\sigma\to \bar{\sigma} + \sigma$, $\chi\to \bar{\chi} + \chi$, and rescale the pion field $\bm{\pi} \to Z \bm{\pi}$
with the wave-function renormalization constant $Z$ for the pseudoscalar fields \cite{Parganlija:2010fz,Janowski:2011gt}. Then, from the terms quadratic
in the fluctuations we read off the pion mass $m_\pi^2 = Z^2(\lambda\bar{\sigma}^2-m^2-2g\bar{\chi})$, and the masses that arise from the mixing of the $\sigma$ 
and $\chi$ fields, 
\be\label{mesonmasses}
m^2_{\pm} = \frac{m_\chi^2+m_\sigma^2}{2}\pm\sqrt{\left(\frac{m_\chi^2-m_\sigma^2}{2}\right)^2 + (2g\bar{\sigma})^2} \, , 
\ee
with $m_\sigma^2 = 3\lambda\bar{\sigma}^2-m^2-2g\bar{\chi}$. Now we notice that, at tree level, the stationarity condition for the condensates
leads to $\epsilon = \bar{\sigma}m_\pi^2/Z^2$. Requiring $\bar{\sigma}=Z f_\pi$ in the vacuum thus determines the parameter $\epsilon=f_\pi m_\pi^2/Z$,
with $m_\pi = 139\,{\rm MeV}$, $f_\pi = 92.4\,{\rm MeV}$, and $Z=1.67$ \cite{Parganlija:2010fz}. We may use the expressions for $m_\sigma$ and $m_\pi$ to 
write
\be \label{lambdamsq}
\lambda = \frac{1}{2(Zf_\pi)^2}\left(m_\sigma^2-\frac{m_\pi^2}{Z^2}\right) \, , \qquad m^2 = \frac{1}{2}\left(m_\sigma^2 - 3\frac{m_\pi^2}{Z^2}\right) 
-\frac{2g^2(Zf_\pi)^2}{m_\chi^2} \, .
\ee
This reparametrizes the remaining constants $m$, $\lambda$,
$m_\chi$, $g$ as functions of $m_\sigma$, $m_\chi$, $g$. 
The model contains 4 more parameters that characterize the interactions between mesons and nucleons (\ref{LIeLSM}), 
$\hat{g}_1$, $\hat{g}_2$, $g_\omega$, $a$. The nucleonic properties 
we need to reproduce are given by the 4 saturation properties (\ref{nEKM}) plus the 2 vacuum masses $m_N = 939\,{\rm MeV}$ and $m_{N^*} = 1535 \, {\rm MeV}$.  
We may now proceed as follows. Again, we can determine $g_\omega$ separately and completely analogously to the Walecka model, using Eq.\ (\ref{stateLSM3}). 
With $a=m_0 m_\chi^2g^{-1}(Zf_\pi)^{-2}$ we can use $m_0$ as a parameter instead of $a$. The advantage of this rescaling is that, if we   
also rescale the condensate $\bar{\chi}=\bar{\chi}'g(Zf_\pi)^{2}m_\chi^{-2}$, the pressure only depends on the ratio $g/m_\chi$, not on 
$g$ and $m_\chi$ separately.
Now we can use Eq.\ (\ref{MNMNs}) to express $\hat{g}_1$ and $\hat{g}_2$ in terms of $m_0$ and the vacuum masses $m_N$, $m_{N^*}$ (and $Z$ and $f_\pi$). 
The remaining parameters $m_\sigma$, $g/m_\chi$ and $m_0$ are determined by solving the following four coupled equations for $m_\sigma$, $g/m_\chi$, $m_0$, 
and $\bar{\chi}'$: the remaining two equations to minimize the free energy, i.e., Eqs.\ (\ref{stateLSM1}), (\ref{stateLSM2}) in the limit $B\to 0$, 
and, as in the Walecka model, the condition that the pressures 
of nuclear matter at the onset and the vacuum are identical, and the condition for the compression modulus. The compression modulus
assumes the same form as given in Eq.\ (\ref{comp}), but with the replacement 
\be 
\frac{\partial^2 U}{\partial M_N^2} \to \frac{\partial^2 U}{\partial M_N^2}-\left(\frac{\partial^2 U}{\partial M_N\partial M_{N^*}}\right)^2
\left/\frac{\partial^2 U}{\partial M_{N^*}^2}\right. \, , 
\ee
because both equations for the condensates (minimization with respect to $\bar{\sigma}$ and $\bar{\chi}$ or equivalently with respect to $M_N$ and $M_{N^*}$) have to 
be taken into account in computing the dependence of $M_N$ on $n$. 

By demanding that $m_+$ that arises from the mixing of the $\sigma$ and $\chi$ fields given in Eq. (\ref{mesonmasses}) is in agreement with the resonance $f_0(1370)$ and keeping $g/m_\chi$ as well as $m_\sigma$ fixed we obtain the values for $g$ and $m_\chi$. This leads to a physical mass $m_-\simeq 715.08\, {\rm MeV}$ for the lighter meson, which is in (rough) accordance with the $f_0(500)$ resonance. Note however that, compared to the 
parameter sets used in Refs.\ \cite{Gallas:2011qp,Heinz:2013hza}, $\sigma$ and $\chi$ have reversed their roles: with our parameter set, the $f_0(500)$ is predominantly 
given by $\sigma$ (= a quark-antiquark state), while $f_0(1370)$ is predominantly given by $\chi$ (= a tetraquark state). We have checked that, given the 
properties of nuclear matter in Eq.\ (\ref{nEKM}), such a role reversal is unavoidable. The main reason 
is our more realistic choice of the effective mass at saturation $M_N=0.8\, m_N$ (while the original parameter sets lead to $M_N \simeq 0.9\, m_N$). Choosing an 
even lower effective mass would make it very difficult for the model in its present form
to reproduce the resonances $f_0(500)$ and $f_0(1370)$ at all. 

The final result of the fitting procedure is summarized in terms of the parameters of the original Lagrangian in Table \ref{table0}.

\begin{table*}[t]
\begin{tabular}{c||c|c|c|c|c|c|c} 
\hline
\multicolumn{7}{c}{Walecka model}
\\[1ex] \hline\hline
 & $\;\;$$m_\sigma\,[{\rm MeV}]$$\;\;$  & $\;\;$$m_\omega\,[{\rm MeV}]$$\;\;$ & $\;\;$$m_N\,[{\rm MeV}]$$\;\;$ & $\;\;$$g_\omega$$\;\;$ & 
$\;\;$$g_\sigma$$\;\;$ & $\;\;$$b$ $\;\;$
& $c$ \\[1ex] \hline
\rule[-1.5ex]{0em}{4ex} 
$\;\;$no-sea approximation $\;\;$ & $\;\;$$550$$\;\;$ & 782 & $\;\;$939$\;\;$ & $\;\;$8.1617$\;\;$ & $\;\;$8.4264$\;\;$ & $\;\;$$8.7788\times 10^{-3}$$\;\;$ & 
$\;\;$$6.8358\times 10^{-3}$$\;\;$ \\[1ex] \hline
$\;\;$including $\Delta\Omega_N$ $\;\;$& $\;\;$$550$$\;\;$ & 782 & $\;\;$939$\;\;$ & $\;\;$8.1617$\;\;$ & 8.5062 & $1.0784\times 10^{-2}$ & $-6.2205\times 10^{-3}$
\\[1ex] \hline
$\;\;$including $\Delta\Omega_N+\Delta\Omega_\sigma$  $\;\;$& $\;\;$$550$$\;\;$ & 782 & $\;\;$939$\;\;$ & $\;\;$8.1617$\;\;$ & 8.1487 & $5.2855\times 10^{-3}$ & 
$ - 2.3611 \times 10^{-2}$  \\[1ex] \hline
\end{tabular}

\vspace{0.2cm}
\begin{tabular}{c|c|c|c|c|c|c|c|c|c} 
\hline
\multicolumn{10}{c}{extended linear sigma model}
\\[1ex] \hline\hline
$\epsilon\,[{\rm MeV}^3]$ & $\;\;$$m\,[{\rm MeV}]$$\;\;$ & $\lambda$ & $\;\;$$g\,[{\rm MeV}]$$\;\;$ & $\;\;$$m_\chi\,[{\rm MeV}]$$\;\;$ & $\;\;$$m_\omega\,[{\rm MeV}]$ $\;\;$
& $\hat{g}_1$ & $\hat{g}_2$ & $g_\omega$ & $a$ 
\\[1ex] \hline
\rule[-1.5ex]{0em}{4ex} 
$\;\;$$1.0690\times 10^6$$\;\;$ & 518.73 & $\;\;$13.950$\;\;$ & 1422.5 & 1310.4 & 782 & $\;\;$10.239$\;\;$ & $\;\;$17.964$\;\;$ & $\;\;$8.1617$\;\;$ & 
$\;\;$29.839$\;\;$ \\[1ex] \hline
\end{tabular}

\caption{Parameters used for the two models. All parameter sets reproduce the density, binding energy, compression modulus and effective mass at saturation
in the absence of a magnetic field 
given in Eq.\ (\ref{nEKM}). In the case of the Walecka model, we discuss the effect of the $B$-independent sea terms in appendix \ref{AppA}, wherefore there 
are three different parameters sets. In the main text we omit the $B$-independent Dirac sea in both models (called "no-sea approximation" in this table). 
}
\label{table0}
\end{table*}

\section{Negligibility of the $B$-independent sea contribution in the Walecka model}
\label{AppA}

In this appendix, we discuss the $B$-independent sea contribution, which we have separated from the $B$-dependent part in Eq.\ (\ref{resultOmNvac}). 
We shall see that its effect on our results is small, thus proving that the utterly dominant effect of the Dirac sea comes from the terms discussed in the
main text. Here we focus on the renormalization of the Walecka 
model \cite{Lee:1974ma,1977AnPhy.108..301C,Serot:1984ey,Serot:1987gi,1988PhLB..208..335G,1989NuPhA.493..521G,glendenningbook}, 
assuming without proof that the conclusions for our results are the same for the extended linear sigma model.
(However, for a study of the chiral phase transition in the extended linear sigma model, which we do not consider here and where the self-consistent nucleon 
masses are allowed to become very small, the validity of the ``no-sea approximation'' is much less clear.) 

We add counterterms to the Lagrangian to all orders up to fourth order in the scalar field $\sigma$, 
i.e., we read the sigma mass $m_\sigma$, and the couplings
$b$, $c$ as bare (cutoff-dependent) quantities, and write them as $m_\sigma^2 = m_{\sigma,r}^2 + \delta m_\sigma^2$, $b = b_r+\delta b$, $c = c_r+\delta c$, 
with renormalized quantities $m_{\sigma,r}^2$, $b_r$, $c_r$. Including also a counterterm linear in $\sigma$, we thus add 
\be
\delta {\cal L} = -\delta a \,m_N^3 (g_\sigma \sigma) - \frac{\delta m_\sigma^2}{2}\sigma^2 - \frac{\delta b}{3} m_N(g_\sigma \sigma)^3
-\frac{\delta c}{4}(g_\sigma \sigma)^4 
\ee
to the Lagrangian. The resulting counterterms in the tree-level potential can be written as
\bea \label{deltaU}
\delta U &=& \left(\delta a +\frac{\delta m_\sigma^2}{2g_\sigma^2 m_N^2}+\frac{\delta b}{3} +\frac{\delta c}{4}\right) m_N^4 
- \left(\delta a +\frac{\delta m_\sigma^2}{g_\sigma^2 m_N^2}+\delta b +\delta c\right) m_N^3 M_N \non[2ex]
&&+\left(\frac{\delta m_\sigma^2}{2g_\sigma^2 m_N^2}+\delta b +\frac{3\delta c}{2}\right) m_N^2 M_N^2 
-\left(\frac{\delta b}{3} +\delta c\right) m_N M_N^3 + \frac{\delta c}{4} M_N^4 \, .
\eea
The relevant cutoff-dependent terms of the free energy can be separated into $B$-independent and $B$-dependent contributions, see Eq.\ (\ref{resultOmNvac}).
The $B$-dependent contributions are discussed in the main text, and here we focus on the $B$-independent part. 
Regularizing the nucleonic part with the proper time method yields (no counterterms added yet) 
\bea
\Omega_N &=& -4\int\frac{d^3{\bf k}}{(2\pi)^3} \left[\epsilon_k+T\sum_{e=\pm}\ln\left(1+e^{-\frac{\epsilon_k-e\mu_*}{T}}\right)\right] \non[2ex]
&=& \frac{\Lambda^4}{8\pi^2}-\frac{M_N^2\Lambda^2}{4\pi^2}+\frac{M_N^4}{8\pi^2}\ln\frac{\Lambda^2}{\ell^2} 
-\frac{M_N^4}{8\pi^2}\left(\gamma-\frac{3}{2}+\ln\frac{M_N^2}{\ell^2}\right) + \Omega_{N,{\rm mat}}\, ,
\eea
where we have introduced the renormalization scale $\ell$, and where $\Omega_{N,{\rm mat}}$ is the matter part of the nucleon contribution [twice the expression 
from Eq.\ (\ref{OmNmat0}) because we have included both isospin components of the nucleons, which are degenerate in our approximation at $qB=0$]. The result
can also be obtained by expanding Eq.\ (\ref{OmegaBeq0}) for large $\Lambda$. In this appendix, where there is no magnetic field, one could have worked with a 
simple momentum cutoff without changing any physical results; only for consistency with the main part we employ the proper time method.
  
The resulting free energy, including cutoff-dependent terms, is valid for all temperatures $T$ and chemical potentials $\mu$. In particular, 
some of the cutoff-dependent terms depend on $T$ and $\mu$ implicitly through the dynamically determined mass $M_N$. Since the theory must be renormalized 
in the vacuum, all counterterms must be independent of $T$ and $\mu$. Consequently, we must require the coefficients in front of each power of $M_N$ to vanish 
separately. The quadratic and quartic contributions from $\Omega_N$ can be cancelled by an appropriate choice of $\delta b$ and $\delta c$. These, in turn, then induce
cutoff-dependent terms of order $M_N$ and $M_N^3$, as can be seen from $\delta U$ in Eq.\ (\ref{deltaU}). Therefore, also $\delta a$ and $\delta m_\sigma^2$ are needed.
One finds that all cutoff-dependencies are cancelled with 
\begin{subequations} \label{counter}
\bea
\delta a &=& \frac{1}{2\pi^2}\ln\frac{\Lambda^2}{\ell^2}-\frac{\Lambda^2}{2\pi^2 m_N^2} +\delta \tilde{a} \, , 
\qquad  \frac{\delta m_\sigma^2}{g_\sigma^2m_N^2} 
= -\frac{3}{2\pi^2}\ln\frac{\Lambda^2}{\ell^2}+ \frac{\Lambda^2}{2\pi^2 m_N^2} +\frac{\delta \tilde{m}^2_\sigma}{g_\sigma^2m_N^2}
\, , \\[2ex]
\delta b &=&  \frac{3}{2\pi^2}\ln\frac{\Lambda^2}{\ell^2}+\delta \tilde{b} \, , \qquad 
\delta c =  -\frac{1}{2\pi^2}\ln\frac{\Lambda^2}{\ell^2}+\delta \tilde{c} \, .
\eea
\end{subequations}
Here we have added finite (cutoff-independent) contributions $\delta \tilde{a}$, $\delta \tilde{m}_\sigma^2$, $\delta \tilde{b}$, $\delta \tilde{c}$.
The free energy now becomes 
\be \label{Omegarenorm}
\Omega = \frac{\Lambda^4}{8\pi^2} -\frac{m_N^2\Lambda^2}{4\pi^2}+\frac{m_N^4}{8\pi^2}\ln\frac{\Lambda^2}{\ell^2} + U +\Delta \Omega_N + \Omega_{N,{\rm mat}}   \, ,
\ee
where the tree-level potential $U$ from Eq.\ (\ref{Uwalecka}) now only contains renormalized quantities $m_{\sigma,r}^2$, $b_r$, $c_r$, and where 
\be \label{DeltaOmN}
\Delta\Omega_N = \delta \tilde{a}\, m_N^3(g_\sigma\bar{\sigma})+\frac{\delta \tilde{m}_\sigma^2}{2}\bar{\sigma}^2+\frac{\delta \tilde{b}}{3}
m_N(g_\sigma\bar{\sigma})^3+\frac{\delta \tilde{c}}{4}(g_\sigma\bar{\sigma})^4 -\frac{M_N^4}{8\pi^2}\ln\frac{M_N^2}{\ell'^2} \, ,
\ee
with the redefined renormalization scale
\be
\ell'^2\equiv \ell^2 e^{3/2-\gamma} \, .
\ee 
The remaining cutoff-dependent terms in Eq.\ (\ref{Omegarenorm}) are constants and do not affect the physics. 
Therefore, they can be dropped by a simple redefinition of the free energy. 

We may require $\Delta \Omega_N$ to contribute to the free energy only to ${\cal O}(\bar{\sigma}^5)$ and higher \cite{Lee:1974ma,1977AnPhy.108..301C,Serot:1987gi}. 
In other words, we may choose the finite parts of the counterterms to cancel the contributions up to fourth order in $\bar{\sigma}$ from the logarithm in 
Eq.\ (\ref{DeltaOmN}). This yields
\bea
\delta \tilde{a} &=& -\frac{1}{4\pi^2}\left(1+2\ln\frac{m_N^2}{\ell'^2}\right) \, , 
\qquad \frac{\delta \tilde{m}^2_\sigma}{g_\sigma^2m_N^2}  =\frac{1}{4\pi^2}\left(7+6\ln\frac{m_N^2}{\ell'^2}\right) \, , \non[2ex] 
\delta \tilde{b} &=&-\frac{1}{4\pi^2}\left(13+6\ln\frac{m_N^2}{\ell'^2}\right) \, , \qquad 
\delta \tilde{c}= \frac{1}{12\pi^2}\left(25+6\ln\frac{m_N^2}{\ell'^2}\right) \, .
\eea
If we require $\bar{\sigma}=0$ (i.e., $M_N=m_N$) to be a solution to the minimization of the free energy with respect to $\bar{\sigma}$, we need to 
choose the renormalization scale $\ell'=m_N$. In this case, the additional free energy from the renormalization terms is
\be 
\Delta\Omega_N = -\frac{1}{4\pi^2}\left[m_N^3(g_\sigma\bar{\sigma})-\frac{7}{2}m_N^2(g_\sigma\bar{\sigma})^2+\frac{13}{3}m_N(g_\sigma\bar{\sigma})^3
-\frac{25}{12}(g_\sigma\bar{\sigma})^4 + M_N^4\ln\frac{M_N}{m_N}\right] \, ,
\ee
and minimizing the free energy with respect to the scalar condensate now yields
\bea \label{statwal1a}
0 = \frac{\partial\Omega}{\partial \bar{\sigma}} = \frac{\partial U}{\partial \bar{\sigma}} + \frac{g_\sigma}{\pi^2}\left[m_N^2(g_\sigma\bar{\sigma}) 
-\frac{5}{2}m_N(g_\sigma \bar{\sigma})^2 +\frac{11}{6}(g_\sigma\bar{\sigma})^3 + M_N^3\ln\frac{M_N}{m_N}\right] 
+ \frac{\partial\Omega_{N,{\rm mat}}}{\partial \bar{\sigma}}\, .
\eea
This equation replaces Eq.\ (\ref{statwal1}) (for $B\to 0$), while the minimization with respect to $\bar{\omega}_0$ (\ref{statwal2}) remains unaltered. 

\begin{figure}[t] 
\begin{center}
\hbox{\includegraphics[width=0.5\textwidth]{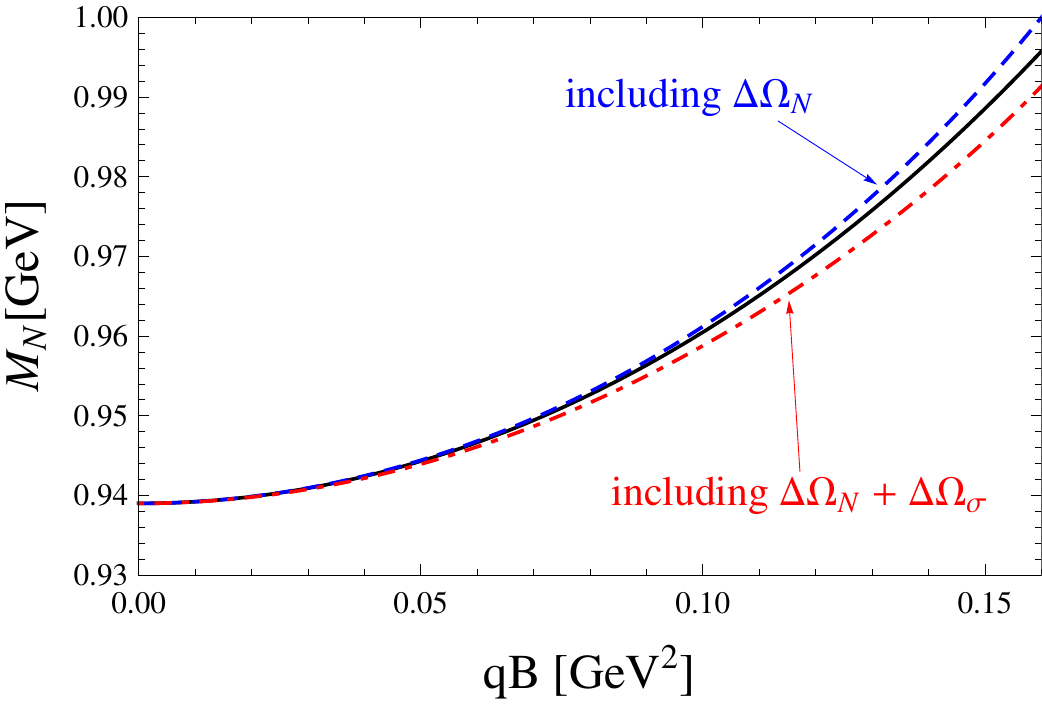}\includegraphics[width=0.5\textwidth]{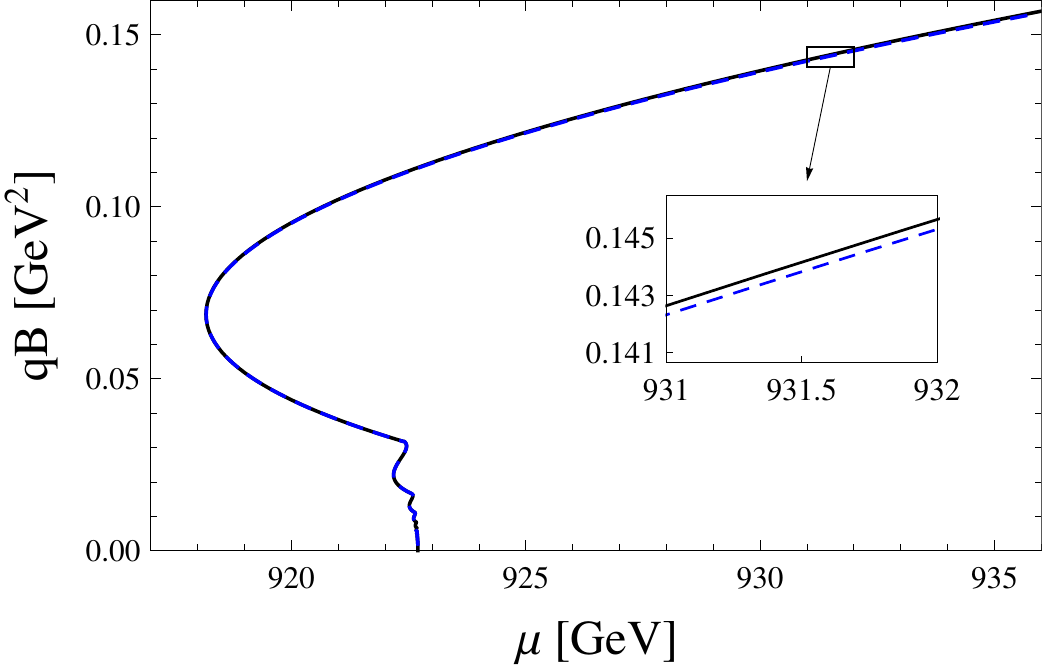}}
\caption{Effect of the $B$-independent sea contributions to the vacuum mass (left) and the onset of nuclear matter (right) in the Walecka model. 
In both panels, the solid line
is the result shown in the main text, i.e., without $B$-independent sea contributions. In the left panel we show the effect of the renormalization of the 
nucleon sector (dashed line) and of the renormalization of both the nucleon and meson sectors (dashed-dotted line). 
The right panel compares the result for the onset with and without renormalizing the nucleon sector, showing that both results are barely distinguishable.}
\label{fig:renorm}
\end{center}
\end{figure}

Due to the additional finite terms in the free energy, the parameters of the model have to be readjusted in order to reproduce the required properties
of nuclear matter. [To this end, we have to replace the tree-level potential $U$ with $U+\Delta\Omega_N$ in the compression modulus (\ref{comp}).]
The resulting parameters are given in Table \ref{table0}. The negative value of $c$ indicates an unbounded tree-level
potential for $\sigma$. This problem is cured if interactions via $\rho$ meson exchange are taken into account, 
see Refs.\ \cite{Serot:1984ey,1988PhLB..208..335G,1989NuPhA.493..521G,glendenningbook} (where also a vacuum contribution $\Delta \Omega_\sigma$ due to 
$\sigma$ loop contributions is included). 

It is now straightforward to extend this renormalization to the case with nonvanishing magnetic field. We can simply treat the $B$-independent 
and $B$-dependent vacuum contributions separately, i.e., we can put together the result from the main part and the result from this appendix,
\be
\Omega = U + \Delta\Omega_N + \frac{B^2}{2}+\Omega_{N,{\rm sea}}  +\Omega_{N,{\rm mat}} \, ,
\ee 
with $\frac{B^2}{2}+\Omega_{N,{\rm sea}}$ given in Eq.\ (\ref{Omrenorm}). In the $B$-independent sea contribution discussed here, a specific choice of $\ell$ 
is needed to proceed. In contrast, in the $B$-dependent sea contribution, $\ell$ only appears in a constant term, and the specific choice of $\ell$ does not matter.
Therefore, any choice (such as $\ell'=m_N$) is compatible with the renormalization discussed in the main text. As argued there, the most 
general choice of the renormalization scale is a combination of the nucleon mass and the magnetic field, and in principle we could include a contribution 
of the magnetic field into $\ell$ here. However, we do not expect the result to change much because we do not consider the regime $|qB|\gg m_N^2$.

In the left panel of Fig.\ \ref{fig:renorm} we show the nucleon mass as a function of the magnetic field. We compare the result obtained after including $\Delta \Omega_N$
with the result used in the main text and shown in Fig.\ \ref{fig:vacmass}. We also show the result obtained after taking the renormalization from the meson loop
into account, by adding \cite{Serot:1984ey,1988PhLB..208..335G,1989NuPhA.493..521G,glendenningbook}     
\be
\Delta \Omega_\sigma = \frac{m_\sigma^4}{(8\pi)^2}\left[(1+\phi_3)^2\ln(1+\phi_3)-\phi_3-\frac{3}{2}\phi_3^2-\frac{1}{3}\phi_1^2(\phi_1+3\phi_2)
+\frac{1}{12}\phi_1^4\right] 
\ee
to the free energy, where
\be
\phi_1\equiv 2 b m_N \frac{g_\sigma^2}{m_\sigma^2}(g_\sigma \bar{\sigma}) \, , \qquad \phi_2 \equiv 3c \frac{g_\sigma^2}{m_\sigma^2}(g_\sigma \bar{\sigma})^2 \, , 
\qquad \phi_3 \equiv \phi_1+\phi_2 \, .
\ee
Including this contribution, the parameters have to be readjusted again, see Table \ref{table0}. 

In the right panel of Fig.\ \ref{fig:renorm} we show the onset with and without $B$-independent sea terms and see that a difference is barely visible. Here we have only
taken into account $\Delta \Omega_N$, not $\Delta \Omega_\sigma$, because our mean-field approximation neglects all meson loops in the medium and thus 
we also neglect them in the vacuum. The main result of this figure and this appendix is that the $B$-{\it independent} sea contributions have no qualitative and 
very little quantitative effect on our results, thus we focus on the $B$-{\it dependent} sea contributions in the main text.

\bibliography{references}

\end{document}